\documentclass[aps,prm,reprint,superscriptaddress]{revtex4-1} 
\usepackage[utf8]{inputenc}
\usepackage{graphicx}
\usepackage{wrapfig}
\usepackage[T1]{fontenc}
\usepackage{float}
\usepackage{xcolor}
 \usepackage[english]{babel}
 \usepackage{amsmath,bm}
 \usepackage{lipsum}
 \usepackage{amsfonts}
 \usepackage{graphicx}
 \usepackage{setspace} 
 \usepackage{graphicx}
 \usepackage{epstopdf}
 \usepackage{dcolumn}
 \usepackage{amsmath}
 \usepackage{epsfig}
 \usepackage{indentfirst}
 \usepackage{psfrag}
 \usepackage{subfigure}
 \usepackage{amssymb}
 \usepackage{color}
 \usepackage{xcolor}
 \usepackage{siunitx}
 \usepackage{graphicx}
 \usepackage{dcolumn}
 \usepackage{bm}
 \usepackage{natbib}
\usepackage{physics}
\usepackage{dcolumn}
\usepackage{bm}
\usepackage{natbib}
\usepackage[backref=none,bookmarksnumbered=true,bookmarks=true,bookmarksopen=true,colorlinks=true,
citecolor=blue,linkcolor=blue,anchorcolor=green,urlcolor=blue,unicode=false]{hyperref}

\usepackage{ulem}[normalem] 

\newcommand{\rev}[1]{{\color{black} #1}}

\begin{document}

\title{Epitaxial growth of gold films on the elemental superconductors \\ V(100), Nb(100) and Nb(110)}

\author{Dongfei Wang}\thanks{These two authors contributed equally}
\email{dfwang@bit.edu.cn}
\affiliation{CIC nanoGUNE-BRTA, 20018 Donostia-San Sebastián, Spain}
\affiliation{School of Physics, Beijing Institute of Technology, Beijing 100081, China}

\author{Katerina Vaxevani}\thanks{These two authors contributed equally}
\affiliation{CIC nanoGUNE-BRTA, 20018 Donostia-San Sebastián, Spain}

\author{Danilo Longo}
\affiliation{CIC nanoGUNE-BRTA, 20018 Donostia-San Sebastián, Spain}

\author{Samuel Kerschbaumer}
\affiliation{Centro de Física de Materiales (CFM-MPC), 20018 Donostia-San Sebastiàn, Spain}

\author{Jon Ortuzar}
\affiliation{CIC nanoGUNE-BRTA, 20018 Donostia-San Sebastián, Spain}

\author{Stefano Trivini}
\affiliation{CIC nanoGUNE-BRTA, 20018 Donostia-San Sebastián, Spain}
\affiliation{Centro de Física de Materiales (CFM-MPC), 20018 Donostia-San Sebastiàn, Spain}

\author{Jingcheng Li}
\affiliation{School of Physics, Sun Yat-sen University, Guangzhou 510275, China}

\author{Maxim Ilyn}
\affiliation{Centro de Física de Materiales (CFM-MPC), 20018 Donostia-San Sebastiàn, Spain}

\author{Celia Rogero}
\affiliation{Centro de Física de Materiales (CFM-MPC), 20018 Donostia-San Sebastiàn, Spain}

\author{Jose Ignacio Pascual}
\email{ji.pascual@nanogune.eu}
\affiliation{CIC nanoGUNE-BRTA, 20018 Donostia-San Sebastián, Spain}
\affiliation{Ikerbasque, Basque Foundation for Science, 48013 Bilbao, Spain}

\date{\today}

\begin{abstract}
Quantum technologies require a new generation of superconducting electronic devices and circuitry. However, the superconducting materials used to construct them are restricted to a class of bulk superconductors.  Gold films grown in contact with superconducting materials can exhibit superconducting correlations through the proximity effect, with various possible implementations in quantum technology. Here, we study the growth of flat Au films on various surfaces of the elemental superconductors vanadium and niobium through a combination of low-temperature scanning tunneling microscopy (STM) and X-ray photoelectron spectroscopy (XPS). In particular, we investigate the growth morphology and composition as a function of temperature and coverage. We find that gold films can grow flat and oxygen-free when annealed to sufficiently high temperatures; however, they are also susceptible to partial intermixing with the substrate elements. Low-temperature scanning tunneling spectroscopy (STS) measurements elucidate the emergence of a proximitized superconducting gap at the gold surface. Additionally, we demonstrate the survival of the magnetic state of FeTPP-Cl molecules on the surface of these gold films, proving that they behave as a good support for probing molecular magnetism. We found that the exchange interaction between the molecular spin and the superconducting condensate can be inferred by the measurement of sub-gap Yu-Shiva-Rusinov states. Our work shows that proximitised Au films constitute a promising platform to explore on-superconducting-surface synthesis and the interaction between superconductivity and magnetism in a large spin-orbit coupling environment.

\end{abstract}

\maketitle
\newpage

\section{Introduction}   

Superconducting materials are nowadays in the spotlight for the accelerating development of quantum information technology \cite{de_leon_materials_2021,devoret_superconducting_2013}. Superconductors not only play a key role in the manufacturing of quantum bits (qubits) but also in probing their quantum state \cite{goppl_coplanar_2008}, minimizing decoherence \cite{place_new_2021} and mediating their inter-coupling \cite{kjaergaard_superconducting_2020}. Taking a step towards this approach, the superconducting proximity effect \cite{proximity_usadel_1970, zhou_density_1998,mcmillan1968tunneling} can serve as a tool to improve the current status quo. First, at the single qubit level, proximitized superconducting substrates are useful in decreasing sources of noise and dissipation, increasing spin relaxation (T1) and coherence (T2) lifetimes \cite{place_new_2021,premkumar_microscopic_2021}. In agreement with this, it was shown that induced superconducting pairing effects in gold (Au) thin films significantly increase T1 of molecular spins lying on their atomically clean surface compared to direct deposition on an elemental superconductor such as Pb \cite{Vaxevani2022extending}. Second, the intrinsic large spin-orbit coupling \cite{tusche2015spin} makes proximitized Au surfaces a very good model system for studying topological superconductivity \cite{qi2011topological,potter2012topological,yan2015topological,chen_2024_signatures}. On these platforms, Majorana bound states, key to topological quantum computation \cite{intro_topological_computation}, are predicted to exist upon further hybridization with magnetic insulators \cite{wei2019superconductivity,manna2020signature,sau2010generic} or by engineering chains of magnetic atoms \cite{nadj2013proposal,peng2015strong,brydon2015topological}. Finally, due to its catalytic properties, the Au surface can serve as a playground to engineer molecules with customised magnetic properties, e.g. graphene nano-ribbons, via on-surface-synthesis (OSS) strategies \cite{cai2010atomically,ruffieux2016surface,liu_proximity-induced_2023}.

The surfaces of most elemental superconductors under investigation generally contain a significant amount of bulk impurities, including oxygen atoms that contribute to surface reconstructions. Such heterogeneous surfaces can hinder OSS processes. Therefore, using Au as a passivating overlayer provides us with an additional motivation to achieve OSS on a superconducting substrate.  In order to advance toward the aforementioned applications of the proximitized Au films, it is necessary to study the growth mechanisms at various temperatures and pinpoint the optimal conditions for achieving atomically flat metallic films. Previous studies of metallic thin films grown on elemental superconductors focus on the chemical composition analysis by X-Ray Photoelectron Spectroscopy (XPS), where an alloy-phase formation was reported above certain annealing temperatures \cite{valla1994photoelectron,lykhach2003thermal,huger2005subsurface,naftel_study_1999,ruckman_growth_1988}. The growth of a thin Au film on superconductors such as V(110) \cite{wei2019superconductivity,wei2016induced} and Nb(110) \cite{beck2023search} has only recently been reported. In such cases, real-space imaging under various growth conditions as well as characterization of the electronic density of states, such as the proximitized superconducting gap, have yet to be explored. 

In this study, we investigate the epitaxial growth of Au thin films on V(100), Nb(110) and Nb(100) single crystals by means of Scanning Tunneling Microscopy (STM). We explore the key parameters for achieving flat thin films on all substrates, such as Au coverage and post-annealing temperatures. Interestingly, for Au on V(100), we find that the post-annealing temperature required to create flat films increases as the Au coverage increases. Additionally, to address the chemical composition of the Au films we performed XPS experiments for various thicknesses and post-annealing temperatures. Overall, we conclude to 3 main points: (1) It is possible to achieve flat Au films on V and Nb substrates. (2) Upon annealing at the temperatures required to obtain flat films, the atoms of the substrate material diffuse and mix with the gold layer. The amount of bulk atoms on the surface depends on the \rev{film} thickness and temperature; (3) \rev{While all the Au films under study are superconducting, only on some surface phases molecules maintain their integrity upon adsorption}. Our results elucidate that flat proximitized Au thin films on top of bulk V and Nb crystals is a promising platform to methodically study the interaction between molecular magnetism and superconductivity.

The paper is organized as follows. In Sec.\ref{Au_growth} we describe the growth of Au films on V and Nb single crystals. 
We include results about the Au film topography and XPS chemical analysis on V(100) and film growth on Nb(110) and Nb(100) surfaces. The electronic properties of Au film measured by STS are presented in Sec.\ref{AuSC}, where we demonstrate that the films exhibit superconductivity due to the proximity effect and characterize their dependence on magnetic field and film thickness. Finally, we evaluate the chemical reactivity of Au films grown on Nb(110) by depositing a magnetic metalloporhyrin molecule, namely FeTPP-Cl. This reveals magnetic coupling between the molecule and the substrate, by the presence of Yu-Shiba-Rusinov \cite{yu1965bound,shiba1968classical,rusinov1969theory} (YSR) subgap resonances.    

\section{Growth of Au on bulk superconductors}\label{Au_growth}

As mentioned before, bulk vanadium and niobium crystals contain a high amount of impurities, with the most common being oxygen, nitrogen, carbon and sulfur. While the impurity concentration in these crystals does not affect their usual bulk superconducting properties, their surface preparation seems to be non-trivial. Typical surface reconstructions appear for V(100), Nb(100) and Nb(110) due to the presence of large amounts of oxygen in the bulk and the high affinity of these transition metals to form stable oxides \cite{huger2005subsurface}.
According to previous reports \cite{kneisel2003surface,shen_oxygen_2007, odobesko2019preparation}, removing the bulk impurities requires very high temperatures, close to the crystal's melting point, which were not always successful \cite{koller2001structure}. Based on the above, we chose to grow gold thin films on top of the flat but reconstructed surfaces of V and Nb. For Ag films grown on top of V and Nb, it has been shown \cite{schneider_proximity_2023,valla_interaction_1995} that pseudomorphic monolayers of Ag are created at the interface, allowing for an ordered Ag structure to develop upon further deposition. However, for Au, the question of how the high miscibility of the underlying transition metals affects the epitaxially grown Au films remains unresolved.

\begin{figure*}[ht]
   \centering
   \includegraphics[width=1\textwidth]{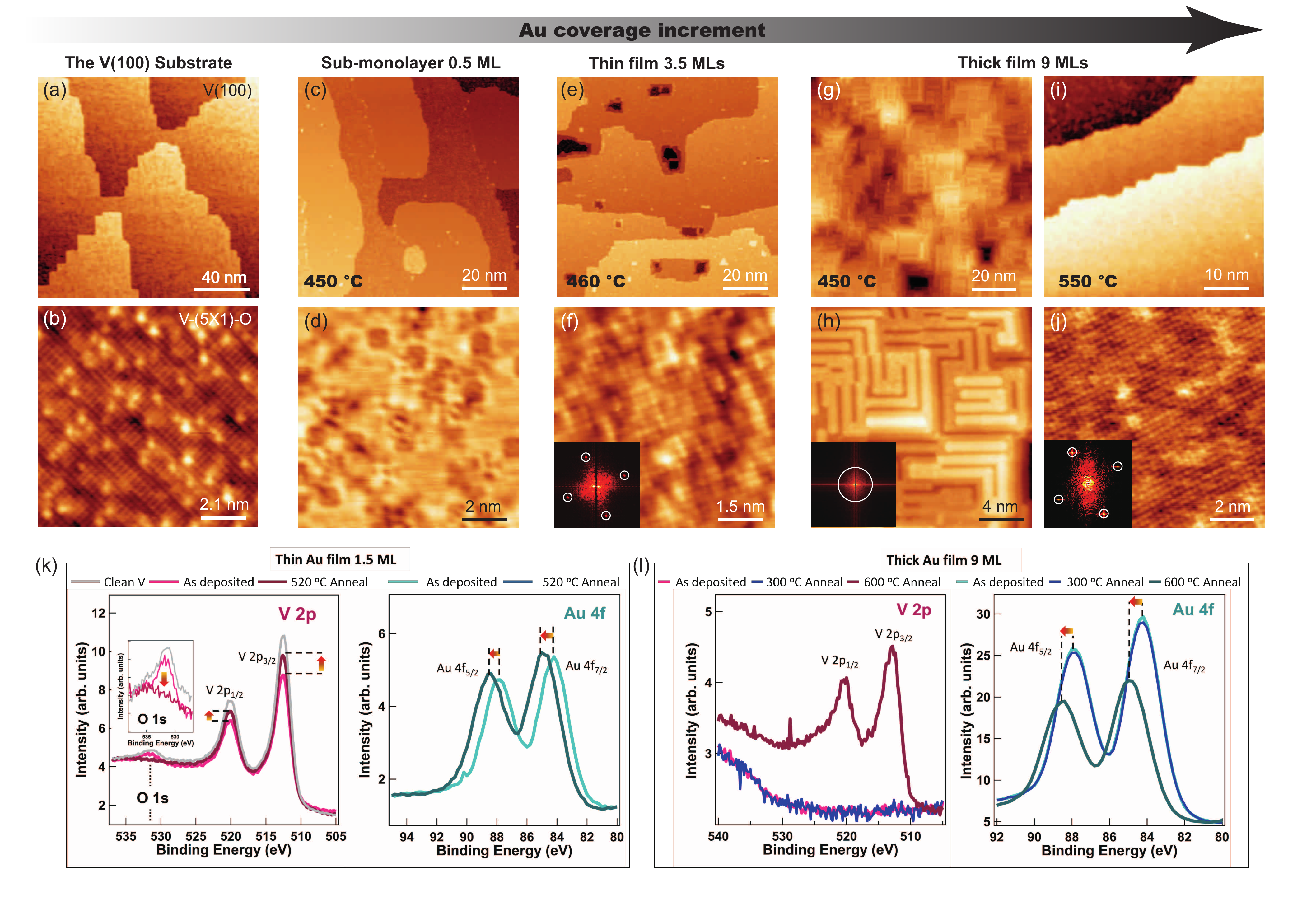}
   \caption{Au films grown on V(100). (a,c,e,g,i) Large scale STM image of clean V(100) surface with V-(\(5\times1\))-O reconstruction (a), with 0.5 ML thick Au film under 450 $\rm^{o}$C post-annealing (c), with 3.5 ML thick Au films under 460 $\rm^{o}$C post-annealing (e), with 9 ML thick Au films under 450 $\rm^{o}$C post-annealing (g) and  with 9 ML thick Au films under 550 $\rm^{o}$C post-annealing (i). (b,d,f,h,j) Small scale STM images corresponding to the Au films shown in (a,c,e,g,f). The insets in (f,h,j) show the Fast-Fourier-Transform (FFT) images. (k,l) XPS data of the V 2p$_{1/2}$, V 2p$_{3/2}$, Au 4f$_{5/2}$ and Au 4f$_{7/2}$ peaks for a 1.5 ML Au film (k) and 9 MLs Au film (l) before and after annealing at a certain temperature. The inset of (k) shows the zoom-in XPS data of the O 1s peak before and after annealing. Tunneling parameters: (a) V=1 V, I=100 pA, (b) V=10 mV, I=50 nA, (c) V=0.5 V, I=400 pA, (d) V=10 mV, I=400 pA, (e) Top V=1 V, I=100 pA, (f) V=80 mV, I=50 nA, (g,h) V=500 mV, I=400 pA, (i) V=1 V, I=400 pA, (j) V=20 mV, I=40 nA.}
   \label{fig:Figure1}
\end{figure*}

We compare different surface terminations of the underling bulk metals in order to pinpoint the diverse advantages of the thin film growth and the resulting surface orientations.  For example, the unreconstructed Au(100) surface is primarily used as a catalyst in on-surface electrochemical reactions, while its affinity with vanadium atoms has been proven useful in studies of single-site catalysts \cite{tempas_2018_redox}.
Therefore, we first attempted to use a V(100) single crystal as a substrate for the growth of gold films. We prepared the vanadium surface via sputtering-annealing cycles followed by several high-temperature flashes (see methods). This resulted in atomically clean and flat terraces of the typical V-(\(5\times1\))-O reconstruction (\autoref{fig:Figure1}a and b).  \autoref{fig:Figure1}(b) shows the atomically-resolved reconstruction whose atomic model is given in Fig. S1(a). The bright protrusions can be identified as oxygen vacancies, consistent with findings reported in previous studies \cite{huang2020quantum,koller2001structure}. The oxygen reconstruction is ubiquitous on this surface, and its removal requires extensive high-temperature cleaning procedures \cite{kralj_hraes_2003,Kralj_1999_stm}.  

We grew gold films directly on the  V-(\(5\times1\))-O surface at room temperature and under ultra-high vacuum (UHV). Under these conditions, no epitaxial growth was observed. Instead, the sample appeared decorated with gold clusters, regardless of the gold coverage (Fig. S2). To induce the re-crystallization of the gold films, we performed subsequent post-annealing steps (more details can be found in Sec. \ref{method}).  
The STM images of Au films with varying thicknesses from 0.5 monolayers (MLs) to 9 MLs are shown in \autoref{fig:Figure1}(c-j). Annealing to $\approx$ 450 $\rm^{o}$C suffices to form flat epitaxial films for lower gold coverages, e.g. for 0.5 MLs  or 3.5 MLs (\autoref{fig:Figure1}(c,e)), but for larger film thicknesses results in recrystallization with large corrugation (e.g. 9 MLs in  \autoref{fig:Figure1}(g,h)).  These thicker films exhibit a 3D structure formed by small terraces decorated with a surface reconstruction with periodicity near 1.3 nm, resembling the 1.4 nm periodic striped surface reconstruction of Au(100)  \cite{yoshimoto2012potential,hammer2014surface,trembulowicz2021100}. However, establishing a direct link with the canonical Au(100) reconstruction is difficult due to the small size of the terraces in the 3D structures. With further annealing to 550 $\rm^{o}$C, the 9 MLs-thick Au films become atomically flat (\autoref{fig:Figure1}(i)). 

In \autoref{fig:Figure1}(d,f,j), we show small-scale STM images of the flat Au films with increasing thicknesses after high-temperature annealing.  For 0.5 ML Au coverage, no atomic-resolved image could be obtained, probably due to disorder. However, for coverages above 2.5 MLs, the surface exhibits periodic atomic structure (\autoref{fig:Figure1}(f,j)). The lattice constant extracted from these images decreases with the coverage, i.e., 306 pm, 292 pm and 286 pm for 2.5 MLs, 3.5 MLs and 9 MLs respectively. These values are in agreement with the gradual transition from a V(100) surface with lattice constant of 303 pm to an Au(100) surface with lattice constant of 288 pm. Although the gold films appear to be flat, the atomically resolved images reveal local surface inhomogeneities, manifested as black depressions and white protrusions. In order to further study the origin of these inhomogenious features, we performed coverage-dependent XPS experiments on Au/V(100) to identify all the atomic species residing in the topmost layers of the Au films.

We first focused on the changes in the XPS peaks for vanadium (V 2p$_{1/2}$ and V 2p$_{3/2}$) and gold (Au 4f$_{5/2}$ and Au 4f$_{7/2}$), specifically in terms of intensity and energy, as a function of temperature and thickness. Deposition of 1.5 ML Au at room temperature leads to a decrease in the intensity for the two characteristic V 2p peaks compared to clean V(100), as it can be seen in the left panel of \autoref{fig:Figure1}(k). At the same time, the Au peaks appear in the spectra at the corresponding binding energies for pure Au (right panel of \autoref{fig:Figure1}(k)). However, upon annealing to 520 $\rm^{o}$C, we observe an increase in the intensity of the vanadium peaks with respect to the spectra taken before annealing the sample (red arrows in \autoref{fig:Figure1}(k) left panel). Similarly, a shift in the position of the Au peaks toward higher binding energies is indicated by the red arrows in \autoref{fig:Figure1}(k) right panel. The latter indicates a chemical change in the coordination of gold present on the surface. Therefore, we can attribute this behaviour to the formation of a Au-V inter-metallic alloy on the surface, which is in agreement with ref.  \cite{valla1994photoelectron}. Our XPS spectra also reveal that the oxygen peak, characteristic of the V-(\(5\times1\))-O reconstruction, disappears upon annealing (inset of \autoref{fig:Figure1}(k)). The carbon peak, which is also a common bulk impurity, follows a similar trend (Fig. S3). This fact discards any implication of oxygen or carbon in the chemical shifts of the Au film. 

We also measured XPS on thicker Au films grown on top of the vanadium surface (\autoref{fig:Figure1}(l)). In this case, the deposition of 9 MLs of gold totally suppresses the substrate signal of V, and low-temperature annealing at 300 $\rm^{o}$C does not induce any changes, contrary to the thin film case. In \autoref{fig:Figure1}(l) left panel, we can observe the two V XPS peaks re-emerge in the spectra after annealing at higher temperatures, i.e. 600 $\rm^{o}$C. This coincides with the temperature required for flattening of the Au films seen in the STM images (\autoref{fig:Figure1}(i)). Accordingly, the Au XPS peaks of the thick film first appear at the value of bulk gold both for the as-deposited film and after annealing to 300 $\rm^{o}$C. Upon annealing at 600 $\rm^{o}$C, the peaks shift again at higher binding energies, similarly to the 1.5 ML film case (\autoref{fig:Figure1}(l) right panel). 

Based on all the aforementioned observations, we can conclude that for the case of Au on V(100) there is a critical temperature, which varies depending on the Au thickness, where flat Au films are formed. An important aftermath of the regular surface is the diffusion of V atoms from the interface towards the surface. Diffused vanadium atoms appear as local inhomogeneities in STM images of this mixed-metallic phase, as shown in \autoref{fig:Figure1}(j). Our XPS analysis confirms that pure gold films can be grown on V(100) after annealing to mild temperatures, though they appear highly corrugated, probably owing to the minor lattice mismatch between Au(100) and V(100).  However, at post-annealing temperatures near 600 $\rm^{o}$C, the Au film becomes flatter as shown in \autoref{fig:Figure1}(i) while an Au-V inter-metallic alloy is created. 

It has been shown before that this intermixing can be avoided by the formation of a pseudomorphic buffer layer between the metal and the superconductor. In the case of Ag on top of Nb(110), its particular surface reconstruction can facilitate the
growth of a bulk-like metallic surface which fully preserves the superconducting gap of
the underlying substrate \cite{schneider_proximity_2023,tomanic_2016_two-band, liu_proximity-induced_2023}. For this reason, we move our attention to the bulk Nb(110) superconducting crystal and we investigate the case of Au and how the preparation parameters, i.e. temperature and
thickness, can affect the growth of the thin films.

\begin{figure*}[ht]
  \centering
   \includegraphics[width=0.7\textwidth]{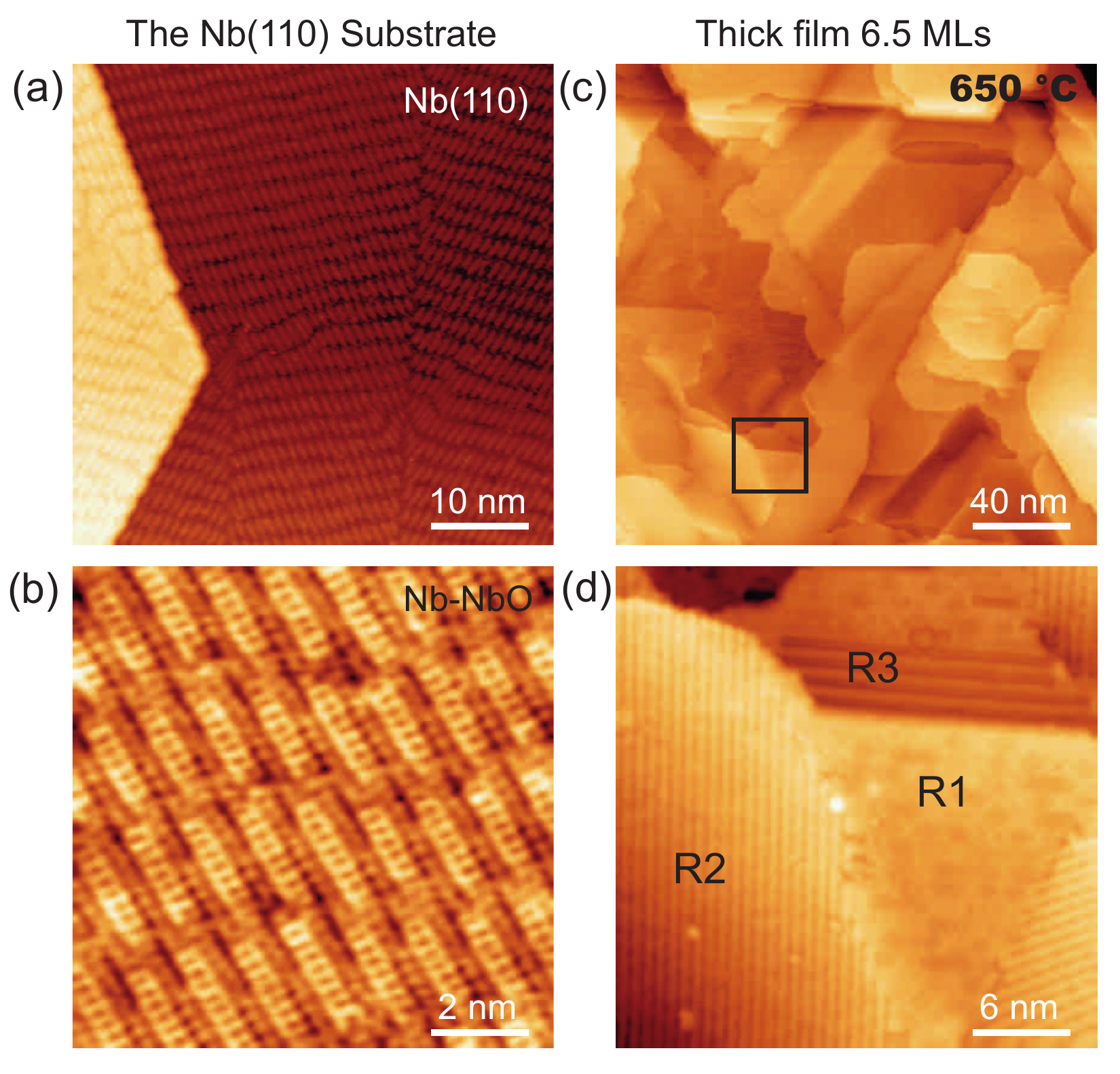}
   \caption{Au films grown on Nb(110). (a,b) Large scale (a) and zoomed in (b) STM images of Nb(110) surface after high temperature treatment. NbO nanocrystals are clearly shown in (b). (c)  Large scale STM images of 6.5 MLs thick Au film at a post-annealing temperature 650 $\rm^{o}$C. (d) A zoomed-in STM image showing the three Au surface terminations R1-R3 with different reconstruction structures. Tunneling parameters: (a,b): V=0.6 V, I=200 pA, (c,d) V=100 mV, I=200 pA}
   \label{fig:Figure2}
\end{figure*}

A (110)-terminated Nb substrate with NbO nanocrystals is formed on the topmost layer which can be seen in \autoref{fig:Figure2}(a,b) and the corresponding model is shown in Fig. S1(b). The appearance of the surface matches with previous reports \cite{zhussupbekov2020oxidation,razinkin2010scanning,arfaoui2004model,berman2023unraveling}. It is worth mentioning that a clean Nb(110) surface without oxygen is possible to obtain by flashing to its melting point \cite{odobesko2019preparation,beck2023search}. Nevertheless, in this experiment, we studied the growth of Au films on the oxygen-reconstructed surface. 

Similar to V(100), we aimed to create a flat, well-formed surface of gold on top of the Nb(110) substrate. We start by depositing sub-monolayer coverage of Au to identify its growth mechanism on Nb(110). As shown in Fig. S2(b,c), deposition at room temperature as well as heat treatment at moderate temperatures did not result in a layer-by-layer (Frank–van der Merwe) growth mode \cite{Theory_levi_1996}, characteristic of a 2D thin film for the formation of the pseudomorphic layer.
Therefore, we continue by evaporating 6.5 MLs of Au on the Nb(110) and post-annealing the sample to higher temperatures similar to the case of thick Au films on V(100), i.e. at 650 $\rm^{o}$C. The resulting 6.5 MLs thick Au films exhibit a more 3D topography with multi-domains as shown in \autoref{fig:Figure2}(c). Nonetheless, the flat Au terraces exceeding 100 nm, visible in the image, are sufficient for our purposes. Three different surface terminations were identified, labeled as R1, R2 and R3 correspondingly in the zoomed-in STM image \autoref{fig:Figure2}(d). The R1 region is flatter compared to the other two and we identify its atomic structure (see also \autoref{fig:Figure4} (c)) with lattice vectors' constant 260 pm and 280 pm which are very close to Au(111) with lattice constant 288.3 pm. However, the R1 surface seems inhomogeneous from the atomically resolved image, very similar to the Au/V(100) case shown previously. Earlier works \cite{huger2005subsurface,ruckman_growth_1988} showed that annealing a gold film on Nb(110) above 427 $\rm^{o}$C induces diffusion of Nb atoms into the film, detected in the XPS signal, as our previous results on V(100). Thus, it is very probable that the flat film we obtained is also alloyed, as indicated by the inhomogeneous atomic resolved image of R1. The R2 and R3 regions both show a one dimensional periodic structure with lattice constant 0.75 nm and 1.13 nm respectively. We attribute region R2 to the Au(110) \(1\times2\) missing row reconstruction with typical lattice constant 0.8 nm and region R3 to its 3-D roughening transition reconstruction with a typical lattice constant 1.2 nm \cite{LandoltBornstein2015:sm_lbs_978-3-662-47736-6_2}.

\begin{figure*}[ht]
  \centering
   \includegraphics[width=1\textwidth]{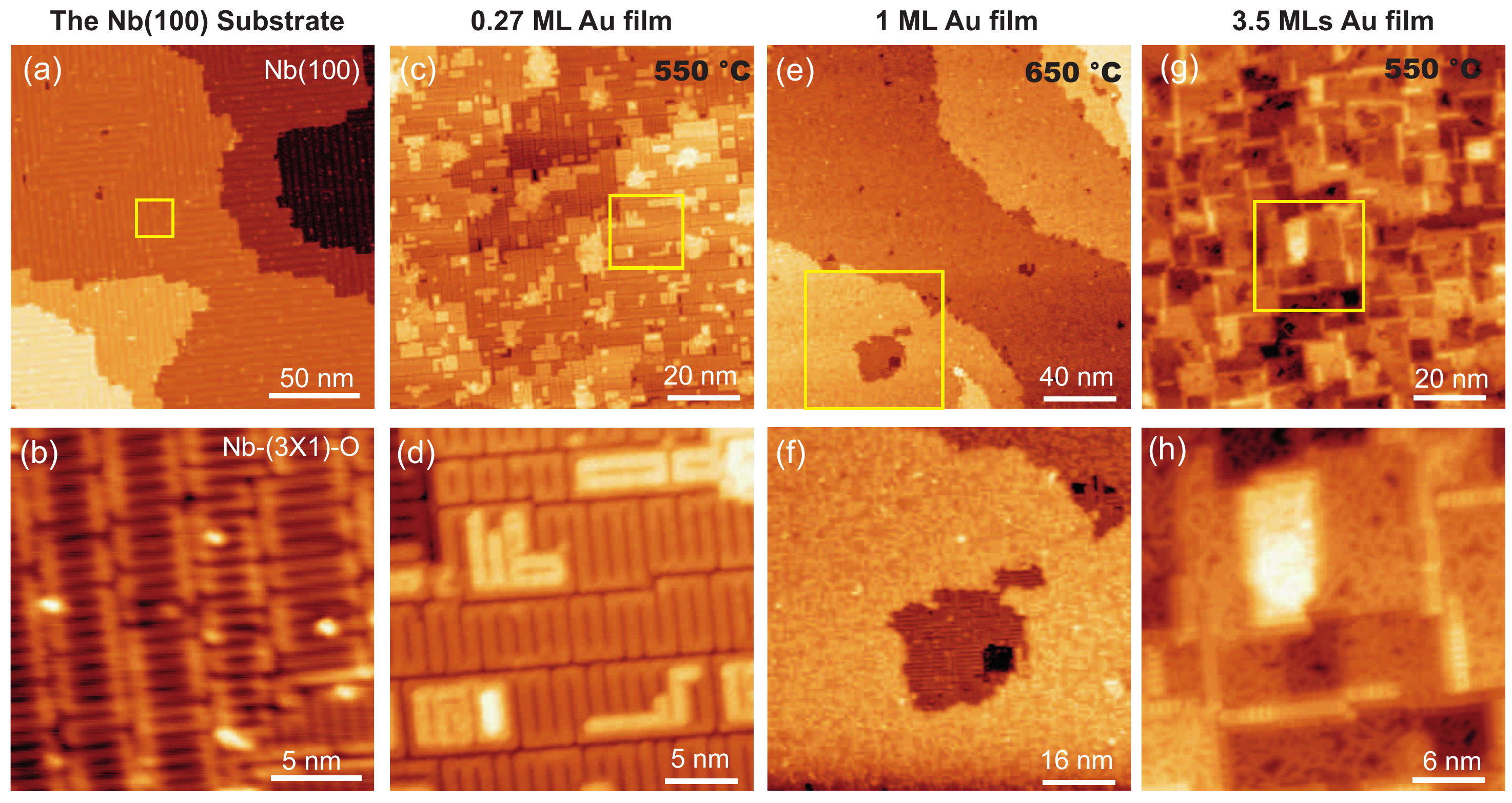}
   \caption{Au films grown on Nb(100). (a,c,e,g) Large scale STM images of Nb(100) surface with Nb-(\(3\times1\))-O surface reconstructions (a), with 0.27 ML Au  after 550 $\rm^{o}$C post-annealing (c), with 1 ML Au after 650 $\rm^{o}$C post-annealing (e) and with 3.5 MLs Au after 550 $\rm^{o}$C post-annealing (g). (b,d,f,h) Zoomed-in STM images of the regions shown in (a,c,e,g). Tunneling parameters: (a,b): V=500 mV, I=100 pA, (c-h) V=340 mV, I=200 pA}
   \label{fig:Figure3}
\end{figure*}

Finally, we present the growth of gold thin films on Nb(100). In order to investigate the effect of the different oxygen-reconstructed surfaces on the growth of the Au thin films, we also tested the deposition on the Nb(100) surface reconstruction (\autoref{fig:Figure3} (a) and (b)). This type of surface topography has been reported before \cite{veit2019oxygen,an2003surface} and was attributed to a Nb-(\(3\times1\))-O reconstruction, with the atomic model shown in Fig. S1(c). 

Au films with thicknesses ranging from sub-monolayer to 3.5 MLs were grown on top of the Nb(100) surface. Fig. S2(d) shows that after deposition of 2ML of Au and annealing at temperatures near 400 $\rm^{o}$C does not result in an organized surface. Similarly to the case of Nb(110), annealing the thin Au films at temperatures above 550 $\rm^{o}$C is necessary to acquire a flat surface. For a sub-monolayer coverage, Au atoms self-organize into wires or rectangular nano-structures following the pattern of Nb-(\(3\times1\))-O reconstruction, as seen in \autoref{fig:Figure3}(c) and (d). This is an indication that the growth mode of the Au film on Nb(100) is of the Volmer–Weber (3D) type even at such low coverage \cite{Theory_levi_1996}. Next, we investigate the growth of near 1 ML coverage of Au and subsequent annealing at 650 $\rm^{o}$C. In \autoref{fig:Figure3}(e) we demonstrate the possibility of obtaining a flat surface upon annealing at this temperature, although it remains rather disordered (\autoref{fig:Figure3}(f)) while no atomic-resolution images could be obtained. For Au films with a thickness of 3.5 MLs and 550 $\rm^{o}$C post-annealing, the surface is characterized by rectangular and three-dimensional facets which are separated by one dimensional domain wall structures, as shown in \autoref{fig:Figure3}(g,h). The growth mode of Au film from sub-ML to 3.5 MLs is similar to the growth reported for iron films grown on W(100), where a transition from pseudomorphic growth to strain relief was observed when increasing the Fe film thickness \cite{wulfhekel2000nano,von2004magnetism}. Additionally, the fact that the growth of Au on Nb(100) is highly directional, especially for the sub-monolayer coverage, can prove very useful for studying the interaction between self-organised atomic chains and superconductivity \cite{nadj2014observation}.

\section{Proximity induced superconductivity in Au films}\label{AuSC}

The Au films grown on bulk superconductors are a promising platform for combining the pairing correlations of the bulk material with the catalytic properties of gold. The proximitization of the thin film depends on the relation between the coherence length and the film thickness and can persist for tens of nanometers in the ballistic regime. \rev{In this regime, quasiparticle scattering is negligible because film thickness is smaller than quasiparticle mean free path, and a hard proximity-induced gap,  i.e. with a total absence of density of states (DoS), is expected on the film, similar to the bulk superconducting gap. } In contrast, if the film has many scattering centers, the superconducting proximitization would follow a diffusive model. In this case, one would expect a reduced gap with a finite density of states, i.e., the so-called dirty superconducting regime \cite{Anderson1959}. 
In this section, we characterize the proximity effect on these films by performing high-resolution differential conductance spectra (dI/dV) at low temperatures (1.5 K). We explore the evolution of the proximitized superconducting gap with respect to the thickness of the Au films up to a few nanometers. Finally, to further study the pair correlations in the gold film, we use a magnetic molecule as a probe system and explore the appearance of Yu-Shiba-Rusinov (YSR) states \cite{yu1965bound,shiba1968classical,rusinov1969theory} in the low energy spectra.

\begin{figure*}[t]
   \centering
   \includegraphics[width=0.7\textwidth]{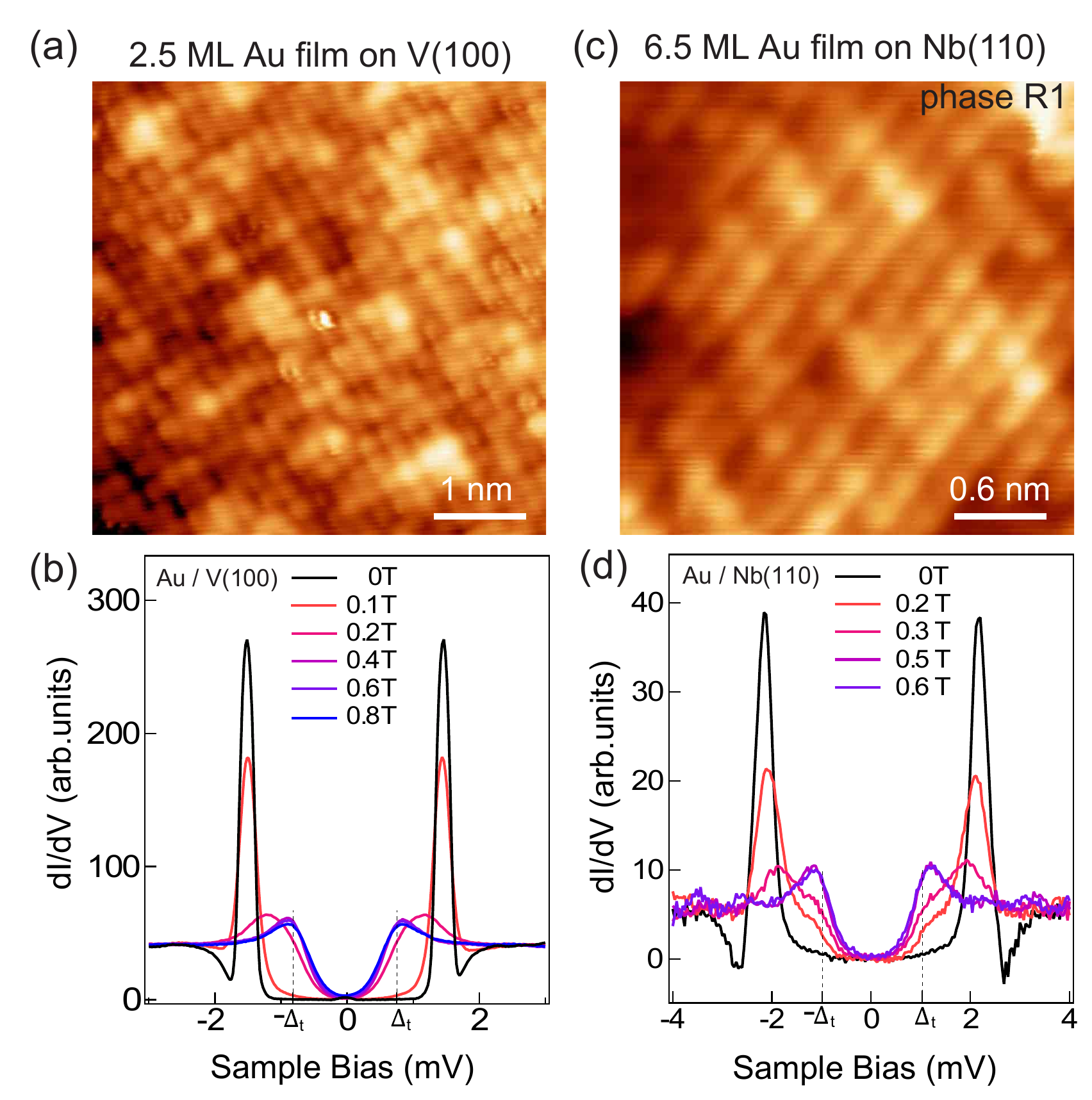}
   \caption{ Superconducting gap measurements for Au films on V(100) and Nb(110). (a) Atomic resolved constant current STM image of 2.5 MLs flat Au film on V(100) prepared with 450 $\rm^{o}$C post-annealing. (b) STS on the surface of 2.5 MLs thick Au film under different magnetic field using a superconducting vanadium tip. (c) Atomic resolved constant height STM image of 6.5 MLs flat Au film on Nb(110) prepared with 650 $\rm^{o}$C post-annealing. (d)  STS on the surface of 6.5 MLs thick Au film under different magnetic field using a superconducting niobium tip. The dI/dV spectra shown in (b) and (d) were taken at 1.3 K. Tunneling parameters: (a) V=5 mV, I=300 pA; (b) V$\rm_{stab}$=5 mV, I$\rm_{stab}$=400 pA; (c) V=0.5 V, Z=-102.618 nm; (d) V$\rm_{stab}$=5 mV, I$\rm_{stab}$=400 pA.}
   \label{fig:Figure4}
\end{figure*}

The study of film growth on V(100) and Nb surfaces revealed that obtaining epitaxial layers requires high-temperature annealing after room-temperature deposition. Consequently, the formation of flat Au terraces is accompanied by a small inter-metallic mixture, where a small amount of V or Nb adatoms can diffuse into the film. Our XPS results confirmed that this is the case for the Au/V(100) system, and their atomically resolved STM images confirm that this appears as a growing amount of atomic sites with different apparent heights (Fig.~\ref{fig:Figure4}a). High-resolution STM images of annealed Au films on Niobium crystals (e.g. as in Fig.~\ref{fig:Figure4}c) also show a growing amount of \textit{defects} attributed to the diffusion of Nb into the film. While this can be problematic for molecular adsorption experiments due to the increased reactivity of the Nb/V-doped Au films, here we analyze whether the superconducting properties of the bulk superconductor persist. 

Figure \ref{fig:Figure4} compares the low-bias dI/dV spectrum of annealed Au films on V(100) and Nb(110). In both cases, we use superconducting tips to enhance the energy resolution and to resolve the sub-gap density of states (DoS).  These tips were obtained by hard indentations into the corresponding bulk superconducting substrate prior to thin-film deposition  \cite{ji2008high,Franke11}. In both cases, the proximitized superconducting gap at zero magnetic field has zero dI/dV signal (it is a hard gap) and it is surrounded by two sharp peaks. These peaks are sub-gap de Gennes-Saint James (dGSJ) resonances, which are Andreev-like quasiparticle states confined between the superconducting/normal metal interface and the vacuum \cite{arnold1978theory,de1963elementary,ortuzar2023theory}. The equation for finding the number of dGSJ resonances that lie inside the superconducting gap reads: 
\begin{equation}
\frac{2L}{\xi} \frac{\epsilon}{\Delta}=n\pi + arccos(\frac{\epsilon}{\Delta})
\end{equation} 
where L is the length of the normal metal, $\xi$ is the coherence length of the superconductor, $\epsilon$ is the energy of the dGSJ resonance and $\Delta$ the value of the superconducting gap\cite{de1963elementary}. Due to the thickness of our films, L<<$\xi$, we expect only one resonant state within the energy range of the superconducting gap for our thickest experimentally grown film. Therefore, we can treat the pair of dGSJ resonances inside the intrinsic superconducting gap of the underlying superconductor as the quasiparticle excitation spectrum of our proximitized substrates\cite{Vaxevani2022extending,trivini2023cooper}.

In both proximitized films, the application of magnetic field causes the disappearance of the dGSJ states and gradual closing of the superconducting gap until a smaller gap, attributed to the STM tip's density of states, appears.  Due to the zero in-gap DoS and low measurement temperatures in our setup, the gap of the tip can only be indirectly observed given that no thermal excitations can take place.  From the set of spectra shown in Figs.~\ref{fig:Figure4}b and Figs.~\ref{fig:Figure4}d, we can estimate that the critical field of the proximitized films amounts to $\sim$0.4 T and $\sim$0.5 T for Au films on V(100) and on Nb(110), respectively. Beyond these values, the spectra reflect the superconducting gap of the tip, measured at $\Delta_t \approx$0.76 meV for the V tip and $\approx$0.96 meV for the Nb tip, as shown in Figs.~\ref{fig:Figure4}b and \ref{fig:Figure4}d (see also Fig.~S4 in supplementary material\cite{SuppMat}). Using these values, we find that the dGSJ states lie at $\pm$0.73 meV and $\pm$1.22 meV inside the V and Nb gaps, corresponding to about 96\% of the bulk gap's value of V(100) (0.76 meV \cite{huang2020quantum}) and 80\% of the bulk gap's value of Nb(110) (1.53 meV \cite{odobesko2019preparation}). We also observed there is no gap value difference between surface terminations R1, R2 and R3 for Au/Nb(110) with similar apparent height (Fig. S5). Hence, we conclude that the thin films of epitaxial Au on V and Nb surfaces, despite being partially doped with material from the bulk supporting substrate, maintain \rev{superconducting properties similar to the bulk. 
These properties are revealed in the spectra by an absolute gap, highlighting the underlying correlated pairs, and a pair of dGSJ peaks at the relative
pairing energies.}

Having resolved the properties of films just a few layers thick, we note that the diffusion of bulk atoms into the Au film might weaken the presence of correlated pairs in thicker films. For crystalline films, we expect that the proximity effect extends for thicknesses of the order of the coherence length scales, that is, tens of nanometers. However, here, the bulk atoms diffused into the Au film can induce new quasiparticle scattering processes that reduce the electronic mean free path and, hence, the extension of the pair correlations. 

\begin{figure*}[t]
\centering
\includegraphics[width=1\textwidth]{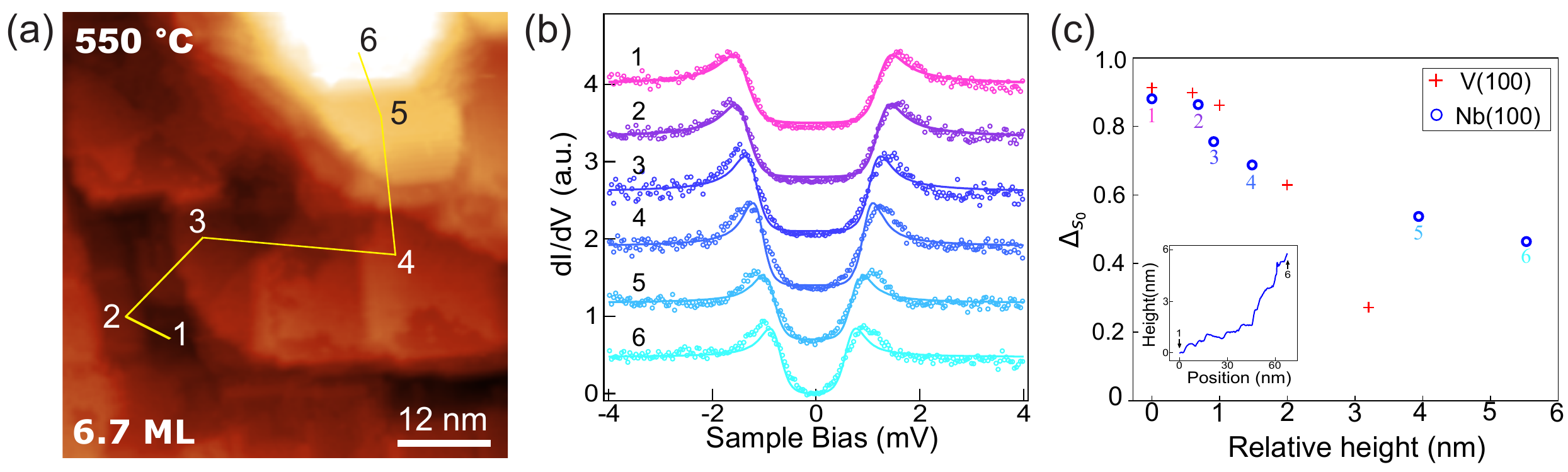}
\caption{Thickness dependent superconducting gap for Au film on Nb(100). (a) Large-scale STM image of 6.7 MLs Au film on Nb(100). The positions where STS were taken are labeled 1-6. (b)  STS measured at positions shown in (a) (circles) with a Dynes’ fit (solid lines). A normal-metal tip was used for spectroscopy.} (c) The dependence of the normalized superconducting gap ($\Delta_{s_0}$) values on the relative Au film apparent height. Blue circles: Gap values measured at positions shown in (a).  Red crosses : Gap values for Au on V(100) taken from ~\cite{Vaxevani2022extending}.  Inset: Line profile along the lines between point 1 and point 6 shown in (a). Tunneling parameters: (a) V=1 V, I=200 pA; (b) V$\rm_{stab}$=5 mV, I$\rm_{stab}$=2 nA. 
\label{fig:Figure5}
\end{figure*}

We have probed the dependence of the proximitized gap on the thickness of the metal films by studying spectra on Au layers with large corrugations. In ~\autoref{fig:Figure5}a, we show an STM image of a 6.7 ML gold film grown on Nb(100) and annealed to 550$^\circ$C. This annealing temperature for such thick layers results in highly corrugated surfaces with terraces about $\sim$ 10 nm wide. In fact, in the region of \autoref{fig:Figure5}a the topographic height changes more than 6 nm, as shown in the height line profile measured between positions 1 and 6 in the inset of \autoref{fig:Figure5}(c). 
Such film corrugation is much larger than the bulk substrate's underneath, which always appear with  extensive flat regions in STM images (e.g. see ~\autoref{fig:Figure3}(a)).
Therefore, we attribute the stacked terraces to regions of the gold film with varying thicknesses. 

This corrugated system permitted us to study the thickness dependence of the superconducting gap by comparing spectra measured on the six positions indicated in  \autoref{fig:Figure5}b using a normal tip, this time. 
At the lowest terrace, the superconducting gap of the substrate is close to the bulk value and decreases monotonously in every higher terrace explored, with a rate of about 10\% of gap reduction per nanometer (blue circles in \autoref{fig:Figure5}c).  
Still, this gap closing with thickness is slower than the one reported for flat annealed Au films on V(100) \cite{Vaxevani2022extending}, shown with crosses in \autoref{fig:Figure5}c. 
The faster decay of proximitized superconductivity in the case of V(100) can be attributed to the higher alloying with bulk atoms for these flatter films.  
Consequently, the fact that superconductivity extends for larger film thickness on the corrugated films on Nb(100) in \autoref{fig:Figure5}c is probably due to less diffusion of Nb atoms into this film, concluded from its three-dimensional shape.  This is further supported by the recent finding \cite{schneider_proximity_2023} that pure silver islands on Nb(110) exhibit a proximitized superconducting gap close to the bulk Nb value for heights of tens of nanometers. From our experiments we conclude that the diffusion of bulk metal atoms into the grown film occurring at higher temperatures unequivocally reduces the ballistic proximity effect in the films.

\subsection*{Proximitized gold films as a support for magnetic molecules}\label{fetpp}

\begin{figure*}[ht]
\centering
\includegraphics[width=1\textwidth]{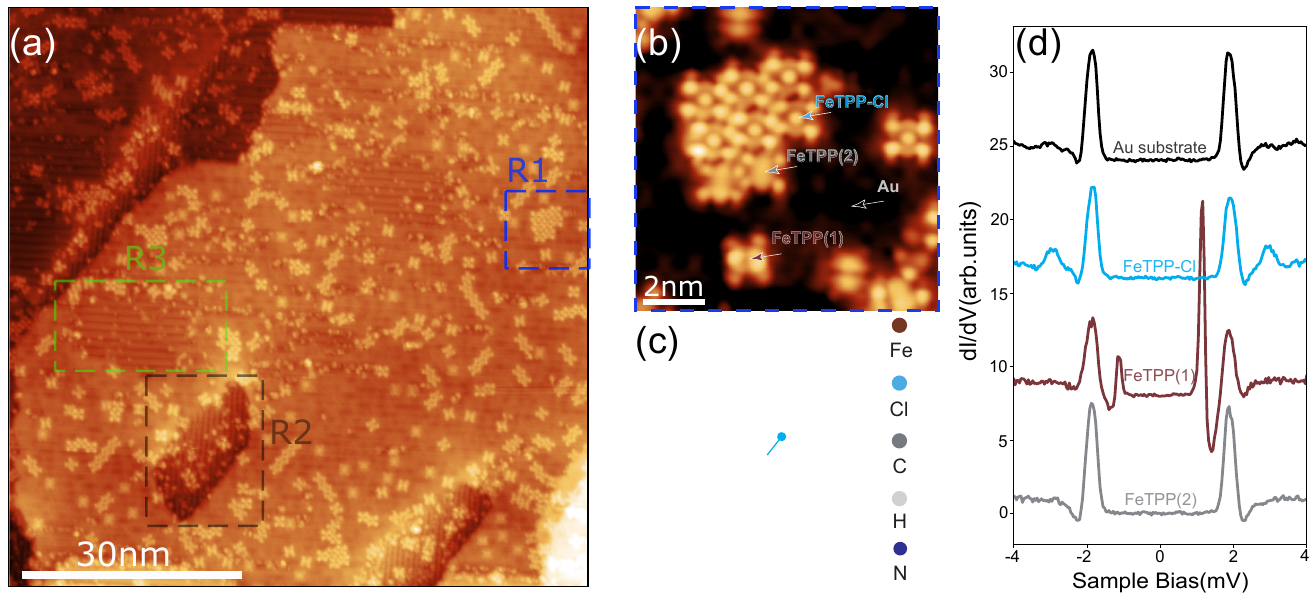}
\caption{Magnetic molecule FeTPP on Au/Nb(110). (a) STM image including all three Au film domains R1 (black square), R2(green square), R3(pink square) on \textcolor{red}{6.5ML of Au on} Nb(110) after sub-ML FeTPP-Cl deposition. (b) Zoomed-in STM image of the FeTPP-Cl and FeTPP molecules in the R1 region in (a). (c)  Structure model of FeTPP-Cl. Zoomed-in STM image of the FeTPP molecule marked by black circle in the R1 region in (b). The arrows indicates where the STS were taken in (d). (d) Low energy range STS taken at the different molecular species indicated by the coloured arrows in (b) measured by a Nb tip with a superconducting gap \(\Delta_{tip}\). The STS taken on the Au film with position away from the molecules is also shown (black curve) for comparison. (a,b) V=100 mV, I=200 pA; (d) V$\rm_{stab}$=5 mV, I=20 pA;}
\label{fig:Figure6}
\end{figure*}

A key concern in molecular nanoscience is whether the magnetic properties of spin-hosting organic species survive in contact with the metal substrate. On several substrates, the spin states in adsorbed molecules disappear due to strong hybridization with the metal, charge transfer, or conformational changes in the magnetic molecules. In our recent publications \cite{Vaxevani2022extending,trivini2023cooper}, we demonstrated that magnetic molecules adsorbed on epitaxial Au films on V(100) maintain their characteristic spin state. Here, we investigate the behavior of the annealed gold films on niobium when hosting magnetic species on its surface. For this purpose, we use the magnetic molecule tetraphenyl porphyrin iron(III) chloride (FeTPP-Cl)  (\autoref{fig:Figure6}(c)). Previous studies of FeTPP-Cl on various substrates revealed that it maintains a S=5/2 spin ground state in the adsorbed state, with a characteristic axial magnetic anisotropy that depends strongly on the molecular conformation \cite{Heinrich13a,Heinrich13b,Heinrich15,Rubio-Verdu18,Vaxevani2022extending}. Additionally, the molecule may undergo dechlorination on the metal substrate, exposing its central Fe metal ion to the substrate, resulting in a characteristic YSR fingerprint \cite{Farinacci20,Rubio-Verdu21a}.

We deposited FeTPP-Cl molecules on annealed Au films on Nb(110) as those shown in \autoref{fig:Figure2}(d). We chose this substrate orientation because the films have a lower corrugation than on the Nb(100) orientation. Additionally, they exhibit different reconstructions with the same superconducting properties (deposition of the same molecular species on annealed gold films on V(100) was reported elsewhere \cite{Vaxevani2022extending, trivini2023cooper}). The adsorption conformation of the molecular species depends on the distinct gold reconstructions, as revealed by STM images like in \autoref{fig:Figure6}(a). 
Region R1 is the area mostly populated with FeTPP-Cl molecules, whereas in regions R2 and R3, the molecules are either decomposed or significantly distorted. These last two phases exhibit characteristic row reconstruction with large atomic corrugation, which usually results in higher reactivity, while phase R1 appears flatter in STM images. We, therefore, remained on the first type of reconstruction for inspecting the molecular magnetic state. 

As indicated in \autoref{fig:Figure6}(b), region R1 hosts various coexisting FeTPP-Cl species, with two types of dechlorinated molecules: one of them (FeTPP(2)) with a characteristic saddle shape and the other (FeTPP(1)) with a chiral structure. The intact species, FeTPP-Cl, is recognized by a central protrusion attributed to the Cl ligand. Both dechlorinated species appear with a nodal plane at the center and are expected to have an integer spin ground state, most probably S=1   \cite{wang2015intramolecularly,liu2017iron,rolf2018visualizing,rolf2019correlation}.  The chiral shape in FeTPP(1) is probably caused by their adsorption configuration with a slight angle with respect to the substrate lattice directions \cite{Mugarza10}.
In \autoref{fig:Figure6}(d), we compare the spectra on the three species with that on the bare substrate. The spectrum on FeTPP-Cl shows two differential conductance (dI/dV) peaks outside the gap attributed to an inelastic excitation from the S=5/2 to the S=3/2 spin state. The inelastic event appears at $\pm 1.0$ meV above the superconducting gap, as a replica of the DoS of the coherence peaks. The corresponding value of the axial magnetic anisotropy constant, $D=$0.5 meV, is smaller on this substrate than for the same molecules on Au(100)/V(100) \cite{Vaxevani2022extending}. While this suggests a greater distortion of FeTPP-Cl molecules on the Au/Nb(110) surface, it also demonstrates that the molecule remains intact. Spectrum on the FeTPP(1) species shows two sub-gap peaks attributed to YSR states on a proximitized gold substrate \cite{ortuzar2023theory}. This spectral fingerprint certifies that the molecule retains its magnetic moment. In contrast, the FeTPP(2) molecule exhibits no magnetic fingerprint, similar to the bare substrate, suggesting that in this case, the molecular spin is quenched by a strong interaction with the substrate. 
These results confirm that the flat proximitized substrates can host magnetic molecules and show many-body states (i.e. YSR states) associated with the interaction between their spin and correlated pairs of the underlying substrate. However, the molecule-surface interaction is larger than on gold single crystal surfaces, probably caused by the intermixing of gold and Nb adatoms at higher annealing temperatures.

\begin{figure*}[ht]
\centering
\includegraphics[width=1\textwidth]{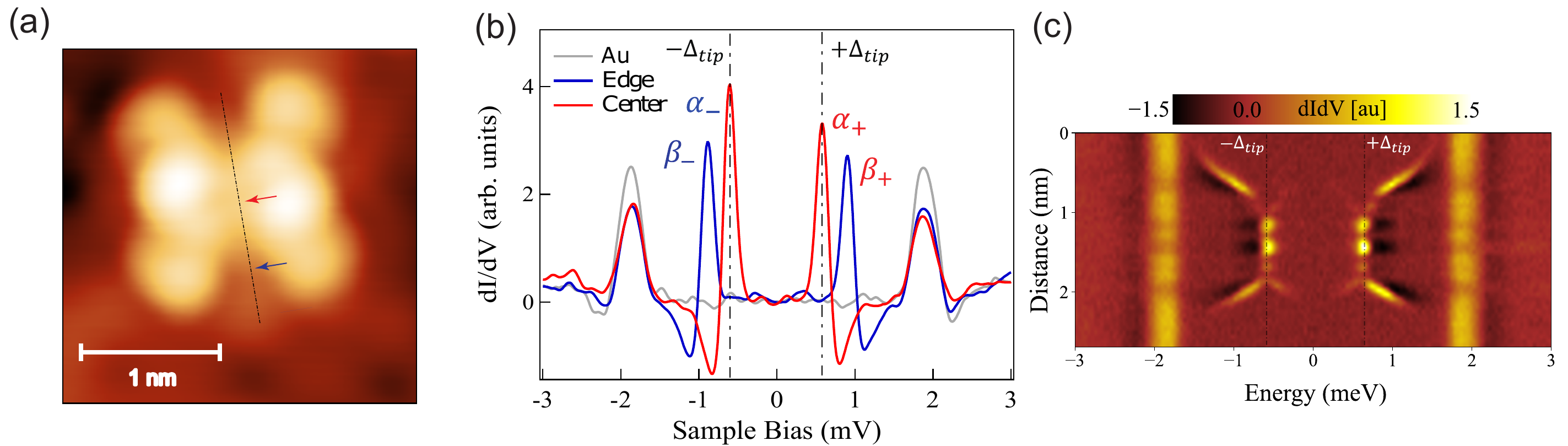}
\caption{(a) Close-up STM image of FeTPP(1). (b) Low energy range STS taken at the center (red curve) and edge (blue curve) of the FeTPP molecule measured by a Nb tip with a superconducting gap \(\Delta_{tip}\). A reference spectrum on the Au film, away from the molecule, is also shown (grey curve) for comparison. YSR bound state with higher and lower intensity are labeled with \(\alpha_-\) (\(\beta_-\)) and \(\alpha_+\) (\(\beta_+\)) correspondingly. (c) Stacked dI/dV spectra taken along the molecular axis indicated in (a). Tunneling parameters: (a) V=10 mV, I=10 pA; (c) V$\rm_{stab}$=5 mV, I=20 pA.}
\label{fig:Figure7}
\end{figure*}

Intriguingly, the slight difference in the adsorption geometry of both dechlorinated species has such an impact on their sub-gap energy spectrum. Such variations could be caused by the different site of the Fe ion on the gold lattice, although we cannot discard that any of the two FeTPP species corresponds to the case of adsorption over a Nb defect. In \autoref{fig:Figure7} we study closely the sub-gap fingerprint of the chiral FeTPP molecule. The pair of YSR resonances appears at the value of the tip gap ($\Delta_{tip}$), which is the zero energy position in the spectral function of the surface (\autoref{fig:Figure7}(b)). Such YSR alignment indicates that chiral FeTPP molecules are at the transition from the strong to weak coupling regime \cite{Franke11}, where the two possible superconducting ground states, a singlet or a doublet, are degenerate. 
However, the YSR resonances appear at different biases at the molecular edges. This is an indication of variations in molecule-substrate exchange caused by the force exerted by the STM tip \cite{farinacci2018tuning,trivini2023cooper}. To confirm this, the dI/dV map of \autoref{fig:Figure7}(c) shows the spectral evolution as the STM tip is moved across the molecular axis indicated in panel (a). \rev{As the tip laterally moves over the molecule, it exerts an increasingly attractive force, resulting in a gradual reduction of the molecular interaction with the substrate \cite{Heinrich15, trivini2023cooper}. This leads to a shift of the YSR resonances towards zero energy (here at $e\Delta_{tip}$ bias). This evolution indicates that the molecule initially lies in a regime of strong coupling with the superconductor, where the ground state consists of a quasiparticle bound to the molecular spin. When the tip is positioned over the center of the molecule, the central Fe atom is maximally pulled away from the surface, and the YSR energy reaches zero (again at $e\Delta_{tip}$ bias), indicating that the system has reached the quantum phase transition to a weaker coupling regime, with a BCS-like pair bound to the molecular spin, although it does not apparently cross this point.} Thus, we can conclude that these species also lie in a strong interaction regime on this metal substrate. 

\section{Conclusion}\label{conclusion}

In conclusion, we have succeeded to grow flat Au films on three different superconducting surfaces, namely V(100), Nb(110) and Nb(100). XPS experiments revealed that the surface of the Au films grown on V(100) is oxygen-free. However, an alloyed Au-V surface is found when post-annealing at temperatures above 500$\rm^{o}$C. In a similar way, flat gold films were grown on Nb(110) after annealing to temperatures above 600$\rm^{o}$C. Here, we found three surface terminations, but only one of them is able to preserve adsorbed molecules intact. Epitaxial films with significant corrugation were also grown on Nb(100), but they exhibited considerable surface disorder.
The Au films on both V and Nb develop a proximitized superconducting gap. For gold films thinner than 10 MLs, the superconducting gap value is greater than half of that of the substrate, which permits to use them as superconducting platforms. Magnetic molecules deposited on top of the proximitized Au films on Nb(110) maintained their spin and molecular structure intact, as we deduced by the analysis of their low energy spectra. Depending on the type of species, we found fingerprints associated to spin excitations or in-gap magnetic bound states. However, in every case, a larger molecule-substrate interaction than on the bare gold substrate was found, indicating that these films are more chemically reactive than the bulk case.  Nevertheless, these results show that superconducting gold films can be grown on various bulk superconductor materials without the need of extensively cleaning their surface and they are a promising platform to study superconductivity-related phenomena in a relatively large spin-orbit interaction background \cite{liu_proximity-induced_2023,liu2022quantum}. 
 
\section{Acknowledgements}

The authors acknowledge financial support from grants PID2019-107338RBC61, PID2022-140845OBC61, PID2020-114252GB-I00, TED2021-130292B-C42 and CEX2020-001038-M funded by MICIU/AEI/10.13039/501100011033 and the European Regional Development Fund (ERDF, A way of making Europe), from grant no. 2023-CIEN-000061-01 funded by the Diputacion de Gipuzkoa, from grant No. IT1591-22 funded by the Basque Government, from the European Union NextGenerationEU/PRTR-C17.I1 as well as by the IKUR Strategy of the Department of Education of the Basque Government, and from the ERC-AdG CONSPIRA (no. 101097693) funded by the European Union (EU). D.W. acknoweledges the JUAN DE LA CIERVA project (Reference:  FJC2020-043831-I) as well as NSFC project No. 12474474.  
K.V. acknowledges enrollment in the doctorate program “Physics of Nanostructures and Advanced Materials” from the advanced polymers and materials, physics, chemistry and technology” department of the Universidad del País Vasco (UPV/EHU). D.L. acknowledges the MSCA project QMOLESR  (project number: 101064332). 
J. O. acknowledges the scholarship PRE\_2021\_1\_0350 from the Basque Government and enrollment in the Physics doctorate program from the Department of Applied Physics of the Universidad del País Vasco (UPV/EHU). 

\section{Methods}\label{method}
Substrate Cleaning: All single crystals, namely Nb(100), Nb(110) and V(100), were firstly sputtered with Ar$^+$ ions 20 minutes with typical ion current 20 uA, following by degassing to 1000 $\rm^{o}$C, in order to remove all oxides formed by exposure to air. Then, the samples were treated with repeated sputtering and annealing up to 1000 $\rm^{o}$C to acquire a flat and clean reconstructed surface. Higher temperature flashes (~1600 $\rm^{o}$C) were employed to achieve longer range flat terraces.
An EFM-3 evaporator loaded with a Mo crucible was used to deposit gold on the cleaned substrates under an ultrahigh vacuum. The Au deposition rate was monitored by measuring residual ion currents and using a quartz microbalance. The final Au coverage was confirmed and calibrated by STM images.  During the deposition, the sample was kept at room temperature and the chamber pressure is better than $1\times10^{-9}$ mbar. After the deposition, the substrate was annealed  and the temperature was monitored with a pyrometer. After the growth, the sample was immediately transferred to the low-temperature STM stage for final inspection. 

STM and STS measurements were performed in ultrahigh vacuum. Spectra were recorded at a temperature of 1.1 K using a lock-in amplifier with a typical frequency of 980 Hz and oscillation amplitude of 0.1 mV. Before switching off the feedback to record the spectra, the tip is stabilized at tunneling conductances on the order of $10^{-4}\times2e^{2}/h$. We used bulk W tip indented on the superconducting substrate to cover them with the bulk superconducting material, vanadium or niobium, to turn them superconducting. The magnetic ﬁelds applied is perpendicular to the sample surface.

The XPS measurements were carried out using a Phoibos-100 electron analyzer (SPECS GmbH), using a non-monochromatic Al K$\alpha$ photon source of 1486.6 eV. The spectra were calibrated to the substrate's main core level (V 2p).


\begin{thebibliography}{81}%
\makeatletter
\providecommand \@ifxundefined [1]{%
 \@ifx{#1\undefined}
}%
\providecommand \@ifnum [1]{%
 \ifnum #1\expandafter \@firstoftwo
 \else \expandafter \@secondoftwo
 \fi
}%
\providecommand \@ifx [1]{%
 \ifx #1\expandafter \@firstoftwo
 \else \expandafter \@secondoftwo
 \fi
}%
\providecommand \natexlab [1]{#1}%
\providecommand \enquote  [1]{``#1''}%
\providecommand \bibnamefont  [1]{#1}%
\providecommand \bibfnamefont [1]{#1}%
\providecommand \citenamefont [1]{#1}%
\providecommand \href@noop [0]{\@secondoftwo}%
\providecommand \href [0]{\begingroup \@sanitize@url \@href}%
\providecommand \@href[1]{\@@startlink{#1}\@@href}%
\providecommand \@@href[1]{\endgroup#1\@@endlink}%
\providecommand \@sanitize@url [0]{\catcode `\\12\catcode `\$12\catcode
  `\&12\catcode `\#12\catcode `\^12\catcode `\_12\catcode `\%12\relax}%
\providecommand \@@startlink[1]{}%
\providecommand \@@endlink[0]{}%
\providecommand \url  [0]{\begingroup\@sanitize@url \@url }%
\providecommand \@url [1]{\endgroup\@href {#1}{\urlprefix }}%
\providecommand \urlprefix  [0]{URL }%
\providecommand \Eprint [0]{\href }%
\providecommand \doibase [0]{http://dx.doi.org/}%
\providecommand \selectlanguage [0]{\@gobble}%
\providecommand \bibinfo  [0]{\@secondoftwo}%
\providecommand \bibfield  [0]{\@secondoftwo}%
\providecommand \translation [1]{[#1]}%
\providecommand \BibitemOpen [0]{}%
\providecommand \bibitemStop [0]{}%
\providecommand \bibitemNoStop [0]{.\EOS\space}%
\providecommand \EOS [0]{\spacefactor3000\relax}%
\providecommand \BibitemShut  [1]{\csname bibitem#1\endcsname}%
\let\auto@bib@innerbib\@empty
\bibitem [{\citenamefont {de~Leon}\ \emph {et~al.}(2021)\citenamefont
  {de~Leon}, \citenamefont {Itoh}, \citenamefont {Kim}, \citenamefont {Mehta},
  \citenamefont {Northup}, \citenamefont {Paik}, \citenamefont {Palmer},
  \citenamefont {Samarth}, \citenamefont {Sangtawesin},\ and\ \citenamefont
  {Steuerman}}]{de_leon_materials_2021}%
  \BibitemOpen
  \bibfield  {author} {\bibinfo {author} {\bibfnamefont {N.~P.}\ \bibnamefont
  {de~Leon}}, \bibinfo {author} {\bibfnamefont {K.~M.}\ \bibnamefont {Itoh}},
  \bibinfo {author} {\bibfnamefont {D.}~\bibnamefont {Kim}}, \bibinfo {author}
  {\bibfnamefont {K.~K.}\ \bibnamefont {Mehta}}, \bibinfo {author}
  {\bibfnamefont {T.~E.}\ \bibnamefont {Northup}}, \bibinfo {author}
  {\bibfnamefont {H.}~\bibnamefont {Paik}}, \bibinfo {author} {\bibfnamefont
  {B.~S.}\ \bibnamefont {Palmer}}, \bibinfo {author} {\bibfnamefont
  {N.}~\bibnamefont {Samarth}}, \bibinfo {author} {\bibfnamefont
  {S.}~\bibnamefont {Sangtawesin}}, \ and\ \bibinfo {author} {\bibfnamefont
  {D.~W.}\ \bibnamefont {Steuerman}},\ }\href@noop {} {\bibfield  {journal}
  {\bibinfo  {journal} {Science}\ }\textbf {\bibinfo {volume} {372}},\ \bibinfo
  {pages} {eabb2823} (\bibinfo {year} {2021})}\BibitemShut {NoStop}%
\bibitem [{\citenamefont {Devoret}\ and\ \citenamefont
  {Schoelkopf}(2013)}]{devoret_superconducting_2013}%
  \BibitemOpen
  \bibfield  {author} {\bibinfo {author} {\bibfnamefont {M.~H.}\ \bibnamefont
  {Devoret}}\ and\ \bibinfo {author} {\bibfnamefont {R.~J.}\ \bibnamefont
  {Schoelkopf}},\ }\href@noop {} {\bibfield  {journal} {\bibinfo  {journal}
  {Science}\ }\textbf {\bibinfo {volume} {339}},\ \bibinfo {pages} {1169}
  (\bibinfo {year} {2013})}\BibitemShut {NoStop}%
\bibitem [{\citenamefont {Göppl}\ \emph {et~al.}(2008)\citenamefont {Göppl},
  \citenamefont {Fragner}, \citenamefont {Baur}, \citenamefont {Bianchetti},
  \citenamefont {Filipp}, \citenamefont {Fink}, \citenamefont {Leek},
  \citenamefont {Puebla}, \citenamefont {Steffen},\ and\ \citenamefont
  {Wallraff}}]{goppl_coplanar_2008}%
  \BibitemOpen
  \bibfield  {author} {\bibinfo {author} {\bibfnamefont {M.}~\bibnamefont
  {Göppl}}, \bibinfo {author} {\bibfnamefont {A.}~\bibnamefont {Fragner}},
  \bibinfo {author} {\bibfnamefont {M.}~\bibnamefont {Baur}}, \bibinfo {author}
  {\bibfnamefont {R.}~\bibnamefont {Bianchetti}}, \bibinfo {author}
  {\bibfnamefont {S.}~\bibnamefont {Filipp}}, \bibinfo {author} {\bibfnamefont
  {J.~M.}\ \bibnamefont {Fink}}, \bibinfo {author} {\bibfnamefont {P.~J.}\
  \bibnamefont {Leek}}, \bibinfo {author} {\bibfnamefont {G.}~\bibnamefont
  {Puebla}}, \bibinfo {author} {\bibfnamefont {L.}~\bibnamefont {Steffen}}, \
  and\ \bibinfo {author} {\bibfnamefont {A.}~\bibnamefont {Wallraff}},\
  }\href@noop {} {\bibfield  {journal} {\bibinfo  {journal} {Journal of Applied
  Physics}\ }\textbf {\bibinfo {volume} {104}},\ \bibinfo {pages} {113904}
  (\bibinfo {year} {2008})}\BibitemShut {NoStop}%
\bibitem [{\citenamefont {Place}\ \emph {et~al.}(2021)\citenamefont {Place},
  \citenamefont {Rodgers}, \citenamefont {Mundada}, \citenamefont {Smitham},
  \citenamefont {Fitzpatrick}, \citenamefont {Leng}, \citenamefont {Premkumar},
  \citenamefont {Bryon}, \citenamefont {Vrajitoarea}, \citenamefont {Sussman},
  \citenamefont {Cheng}, \citenamefont {Madhavan}, \citenamefont {Babla},
  \citenamefont {Le}, \citenamefont {Gang}, \citenamefont {Jäck},
  \citenamefont {Gyenis}, \citenamefont {Yao}, \citenamefont {Cava},
  \citenamefont {de~Leon},\ and\ \citenamefont {Houck}}]{place_new_2021}%
  \BibitemOpen
  \bibfield  {author} {\bibinfo {author} {\bibfnamefont {A.~P.~M.}\
  \bibnamefont {Place}}, \bibinfo {author} {\bibfnamefont {L.~V.~H.}\
  \bibnamefont {Rodgers}}, \bibinfo {author} {\bibfnamefont {P.}~\bibnamefont
  {Mundada}}, \bibinfo {author} {\bibfnamefont {B.~M.}\ \bibnamefont
  {Smitham}}, \bibinfo {author} {\bibfnamefont {M.}~\bibnamefont
  {Fitzpatrick}}, \bibinfo {author} {\bibfnamefont {Z.}~\bibnamefont {Leng}},
  \bibinfo {author} {\bibfnamefont {A.}~\bibnamefont {Premkumar}}, \bibinfo
  {author} {\bibfnamefont {J.}~\bibnamefont {Bryon}}, \bibinfo {author}
  {\bibfnamefont {A.}~\bibnamefont {Vrajitoarea}}, \bibinfo {author}
  {\bibfnamefont {S.}~\bibnamefont {Sussman}}, \bibinfo {author} {\bibfnamefont
  {G.}~\bibnamefont {Cheng}}, \bibinfo {author} {\bibfnamefont
  {T.}~\bibnamefont {Madhavan}}, \bibinfo {author} {\bibfnamefont {H.~K.}\
  \bibnamefont {Babla}}, \bibinfo {author} {\bibfnamefont {X.~H.}\ \bibnamefont
  {Le}}, \bibinfo {author} {\bibfnamefont {Y.}~\bibnamefont {Gang}}, \bibinfo
  {author} {\bibfnamefont {B.}~\bibnamefont {Jäck}}, \bibinfo {author}
  {\bibfnamefont {A.}~\bibnamefont {Gyenis}}, \bibinfo {author} {\bibfnamefont
  {N.}~\bibnamefont {Yao}}, \bibinfo {author} {\bibfnamefont {R.~J.}\
  \bibnamefont {Cava}}, \bibinfo {author} {\bibfnamefont {N.~P.}\ \bibnamefont
  {de~Leon}}, \ and\ \bibinfo {author} {\bibfnamefont {A.~A.}\ \bibnamefont
  {Houck}},\ }\href@noop {} {\bibfield  {journal} {\bibinfo  {journal} {Nature
  Communications}\ }\textbf {\bibinfo {volume} {12}},\ \bibinfo {pages} {1779}
  (\bibinfo {year} {2021})}\BibitemShut {NoStop}%
\bibitem [{\citenamefont {Kjaergaard}\ \emph {et~al.}(2020)\citenamefont
  {Kjaergaard}, \citenamefont {Schwartz}, \citenamefont {Braumüller},
  \citenamefont {Krantz}, \citenamefont {Wang}, \citenamefont {Gustavsson},\
  and\ \citenamefont {Oliver}}]{kjaergaard_superconducting_2020}%
  \BibitemOpen
  \bibfield  {author} {\bibinfo {author} {\bibfnamefont {M.}~\bibnamefont
  {Kjaergaard}}, \bibinfo {author} {\bibfnamefont {M.~E.}\ \bibnamefont
  {Schwartz}}, \bibinfo {author} {\bibfnamefont {J.}~\bibnamefont
  {Braumüller}}, \bibinfo {author} {\bibfnamefont {P.}~\bibnamefont {Krantz}},
  \bibinfo {author} {\bibfnamefont {J.~I.-J.}\ \bibnamefont {Wang}}, \bibinfo
  {author} {\bibfnamefont {S.}~\bibnamefont {Gustavsson}}, \ and\ \bibinfo
  {author} {\bibfnamefont {W.~D.}\ \bibnamefont {Oliver}},\ }\href@noop {}
  {\bibfield  {journal} {\bibinfo  {journal} {Annual Review of Condensed Matter
  Physics}\ }\textbf {\bibinfo {volume} {11}},\ \bibinfo {pages} {369}
  (\bibinfo {year} {2020})}\BibitemShut {NoStop}%
\bibitem [{\citenamefont {Usadel}(1970)}]{proximity_usadel_1970}%
  \BibitemOpen
  \bibfield  {author} {\bibinfo {author} {\bibfnamefont {K.~D.}\ \bibnamefont
  {Usadel}},\ }\href@noop {} {\bibfield  {journal} {\bibinfo  {journal} {Phys.
  Rev. Lett.}\ }\textbf {\bibinfo {volume} {25}},\ \bibinfo {pages} {507}
  (\bibinfo {year} {1970})}\BibitemShut {NoStop}%
\bibitem [{\citenamefont {Zhou}\ \emph {et~al.}(1998)\citenamefont {Zhou},
  \citenamefont {Charlat}, \citenamefont {Spivak},\ and\ \citenamefont
  {Pannetier}}]{zhou_density_1998}%
  \BibitemOpen
  \bibfield  {author} {\bibinfo {author} {\bibfnamefont {F.}~\bibnamefont
  {Zhou}}, \bibinfo {author} {\bibfnamefont {P.}~\bibnamefont {Charlat}},
  \bibinfo {author} {\bibfnamefont {B.}~\bibnamefont {Spivak}}, \ and\ \bibinfo
  {author} {\bibfnamefont {B.}~\bibnamefont {Pannetier}},\ }\href@noop {}
  {\bibfield  {journal} {\bibinfo  {journal} {Journal of Low Temperature
  Physics}\ }\textbf {\bibinfo {volume} {110}},\ \bibinfo {pages} {841}
  (\bibinfo {year} {1998})}\BibitemShut {NoStop}%
\bibitem [{\citenamefont {McMillan}(1968)}]{mcmillan1968tunneling}%
  \BibitemOpen
  \bibfield  {author} {\bibinfo {author} {\bibfnamefont {W.}~\bibnamefont
  {McMillan}},\ }\href@noop {} {\bibfield  {journal} {\bibinfo  {journal}
  {Physical Review}\ }\textbf {\bibinfo {volume} {175}},\ \bibinfo {pages}
  {537} (\bibinfo {year} {1968})}\BibitemShut {NoStop}%
\bibitem [{\citenamefont {Premkumar}\ \emph {et~al.}(2021)\citenamefont
  {Premkumar}, \citenamefont {Weiland}, \citenamefont {Hwang}, \citenamefont
  {Jäck}, \citenamefont {Place}, \citenamefont {Waluyo}, \citenamefont {Hunt},
  \citenamefont {Bisogni}, \citenamefont {Pelliciari}, \citenamefont {Barbour},
  \citenamefont {Miller}, \citenamefont {Russo}, \citenamefont {Camino},
  \citenamefont {Kisslinger}, \citenamefont {Tong}, \citenamefont {Hybertsen},
  \citenamefont {Houck},\ and\ \citenamefont
  {Jarrige}}]{premkumar_microscopic_2021}%
  \BibitemOpen
  \bibfield  {author} {\bibinfo {author} {\bibfnamefont {A.}~\bibnamefont
  {Premkumar}}, \bibinfo {author} {\bibfnamefont {C.}~\bibnamefont {Weiland}},
  \bibinfo {author} {\bibfnamefont {S.}~\bibnamefont {Hwang}}, \bibinfo
  {author} {\bibfnamefont {B.}~\bibnamefont {Jäck}}, \bibinfo {author}
  {\bibfnamefont {A.~P.~M.}\ \bibnamefont {Place}}, \bibinfo {author}
  {\bibfnamefont {I.}~\bibnamefont {Waluyo}}, \bibinfo {author} {\bibfnamefont
  {A.}~\bibnamefont {Hunt}}, \bibinfo {author} {\bibfnamefont {V.}~\bibnamefont
  {Bisogni}}, \bibinfo {author} {\bibfnamefont {J.}~\bibnamefont {Pelliciari}},
  \bibinfo {author} {\bibfnamefont {A.}~\bibnamefont {Barbour}}, \bibinfo
  {author} {\bibfnamefont {M.~S.}\ \bibnamefont {Miller}}, \bibinfo {author}
  {\bibfnamefont {P.}~\bibnamefont {Russo}}, \bibinfo {author} {\bibfnamefont
  {F.}~\bibnamefont {Camino}}, \bibinfo {author} {\bibfnamefont
  {K.}~\bibnamefont {Kisslinger}}, \bibinfo {author} {\bibfnamefont
  {X.}~\bibnamefont {Tong}}, \bibinfo {author} {\bibfnamefont {M.~S.}\
  \bibnamefont {Hybertsen}}, \bibinfo {author} {\bibfnamefont {A.~A.}\
  \bibnamefont {Houck}}, \ and\ \bibinfo {author} {\bibfnamefont
  {I.}~\bibnamefont {Jarrige}},\ }\href@noop {} {\bibfield  {journal} {\bibinfo
   {journal} {Communications Materials}\ }\textbf {\bibinfo {volume} {2}},\
  \bibinfo {pages} {72} (\bibinfo {year} {2021})}\BibitemShut {NoStop}%
\bibitem [{\citenamefont {Vaxevani}\ \emph {et~al.}(2022)\citenamefont
  {Vaxevani}, \citenamefont {Li}, \citenamefont {Trivini}, \citenamefont
  {Ortuzar}, \citenamefont {Longo}, \citenamefont {Wang},\ and\ \citenamefont
  {Pascual}}]{Vaxevani2022extending}%
  \BibitemOpen
  \bibfield  {author} {\bibinfo {author} {\bibfnamefont {K.}~\bibnamefont
  {Vaxevani}}, \bibinfo {author} {\bibfnamefont {J.}~\bibnamefont {Li}},
  \bibinfo {author} {\bibfnamefont {S.}~\bibnamefont {Trivini}}, \bibinfo
  {author} {\bibfnamefont {J.}~\bibnamefont {Ortuzar}}, \bibinfo {author}
  {\bibfnamefont {D.}~\bibnamefont {Longo}}, \bibinfo {author} {\bibfnamefont
  {D.}~\bibnamefont {Wang}}, \ and\ \bibinfo {author} {\bibfnamefont {J.~I.}\
  \bibnamefont {Pascual}},\ }\href@noop {} {\bibfield  {journal} {\bibinfo
  {journal} {Nano Letters}\ }\textbf {\bibinfo {volume} {22}},\ \bibinfo
  {pages} {6075} (\bibinfo {year} {2022})}\BibitemShut {NoStop}%
\bibitem [{\citenamefont {Tusche}\ \emph {et~al.}(2015)\citenamefont {Tusche},
  \citenamefont {Krasyuk},\ and\ \citenamefont {Kirschner}}]{tusche2015spin}%
  \BibitemOpen
  \bibfield  {author} {\bibinfo {author} {\bibfnamefont {C.}~\bibnamefont
  {Tusche}}, \bibinfo {author} {\bibfnamefont {A.}~\bibnamefont {Krasyuk}}, \
  and\ \bibinfo {author} {\bibfnamefont {J.}~\bibnamefont {Kirschner}},\
  }\href@noop {} {\bibfield  {journal} {\bibinfo  {journal} {Ultramicroscopy}\
  }\textbf {\bibinfo {volume} {159}},\ \bibinfo {pages} {520} (\bibinfo {year}
  {2015})}\BibitemShut {NoStop}%
\bibitem [{\citenamefont {Qi}\ and\ \citenamefont
  {Zhang}(2011)}]{qi2011topological}%
  \BibitemOpen
  \bibfield  {author} {\bibinfo {author} {\bibfnamefont {X.-L.}\ \bibnamefont
  {Qi}}\ and\ \bibinfo {author} {\bibfnamefont {S.-C.}\ \bibnamefont {Zhang}},\
  }\href@noop {} {\bibfield  {journal} {\bibinfo  {journal} {Reviews of Modern
  Physics}\ }\textbf {\bibinfo {volume} {83}},\ \bibinfo {pages} {1057}
  (\bibinfo {year} {2011})}\BibitemShut {NoStop}%
\bibitem [{\citenamefont {Potter}\ and\ \citenamefont
  {Lee}(2012)}]{potter2012topological}%
  \BibitemOpen
  \bibfield  {author} {\bibinfo {author} {\bibfnamefont {A.~C.}\ \bibnamefont
  {Potter}}\ and\ \bibinfo {author} {\bibfnamefont {P.~A.}\ \bibnamefont
  {Lee}},\ }\href@noop {} {\bibfield  {journal} {\bibinfo  {journal} {Physical
  Review B}\ }\textbf {\bibinfo {volume} {85}},\ \bibinfo {pages} {094516}
  (\bibinfo {year} {2012})}\BibitemShut {NoStop}%
\bibitem [{\citenamefont {Yan}\ \emph {et~al.}(2015)\citenamefont {Yan},
  \citenamefont {Stadtm{\"u}ller}, \citenamefont {Haag}, \citenamefont
  {Jakobs}, \citenamefont {Seidel}, \citenamefont {Jungkenn}, \citenamefont
  {Mathias}, \citenamefont {Cinchetti}, \citenamefont {Aeschlimann},\ and\
  \citenamefont {Felser}}]{yan2015topological}%
  \BibitemOpen
  \bibfield  {author} {\bibinfo {author} {\bibfnamefont {B.}~\bibnamefont
  {Yan}}, \bibinfo {author} {\bibfnamefont {B.}~\bibnamefont
  {Stadtm{\"u}ller}}, \bibinfo {author} {\bibfnamefont {N.}~\bibnamefont
  {Haag}}, \bibinfo {author} {\bibfnamefont {S.}~\bibnamefont {Jakobs}},
  \bibinfo {author} {\bibfnamefont {J.}~\bibnamefont {Seidel}}, \bibinfo
  {author} {\bibfnamefont {D.}~\bibnamefont {Jungkenn}}, \bibinfo {author}
  {\bibfnamefont {S.}~\bibnamefont {Mathias}}, \bibinfo {author} {\bibfnamefont
  {M.}~\bibnamefont {Cinchetti}}, \bibinfo {author} {\bibfnamefont
  {M.}~\bibnamefont {Aeschlimann}}, \ and\ \bibinfo {author} {\bibfnamefont
  {C.}~\bibnamefont {Felser}},\ }\href@noop {} {\bibfield  {journal} {\bibinfo
  {journal} {Nature communications}\ }\textbf {\bibinfo {volume} {6}},\
  \bibinfo {pages} {1} (\bibinfo {year} {2015})}\BibitemShut {NoStop}%
\bibitem [{\citenamefont {C}\ \emph {et~al.}(2024)\citenamefont {C},
  \citenamefont {Tran}, \citenamefont {McFadden}, \citenamefont {Simmonds},
  \citenamefont {Saito}, \citenamefont {Chu}, \citenamefont {Morales},
  \citenamefont {Suezaki}, \citenamefont {Hou}, \citenamefont {Aumentado},
  \citenamefont {Lee}, \citenamefont {Moodera},\ and\ \citenamefont
  {Wei}}]{chen_2024_signatures}%
  \BibitemOpen
  \bibfield  {author} {\bibinfo {author} {\bibfnamefont {C.}~\bibnamefont {C}},
  \bibinfo {author} {\bibfnamefont {J.}~\bibnamefont {Tran}}, \bibinfo {author}
  {\bibfnamefont {A.}~\bibnamefont {McFadden}}, \bibinfo {author}
  {\bibfnamefont {R.}~\bibnamefont {Simmonds}}, \bibinfo {author}
  {\bibfnamefont {K.}~\bibnamefont {Saito}}, \bibinfo {author} {\bibfnamefont
  {E.-D.}\ \bibnamefont {Chu}}, \bibinfo {author} {\bibfnamefont
  {D.}~\bibnamefont {Morales}}, \bibinfo {author} {\bibfnamefont
  {V.}~\bibnamefont {Suezaki}}, \bibinfo {author} {\bibfnamefont
  {Y.}~\bibnamefont {Hou}}, \bibinfo {author} {\bibfnamefont {J.}~\bibnamefont
  {Aumentado}}, \bibinfo {author} {\bibfnamefont {P.~A.}\ \bibnamefont {Lee}},
  \bibinfo {author} {\bibfnamefont {J.~S.}\ \bibnamefont {Moodera}}, \ and\
  \bibinfo {author} {\bibfnamefont {P.}~\bibnamefont {Wei}},\ }\href@noop {}
  {\bibfield  {journal} {\bibinfo  {journal} {Science Advances}\ }\textbf
  {\bibinfo {volume} {10}},\ \bibinfo {pages} {eado4875} (\bibinfo {year}
  {2024})}\BibitemShut {NoStop}%
\bibitem [{\citenamefont {Lahtinen}\ and\ \citenamefont
  {Pachos}(2017)}]{intro_topological_computation}%
  \BibitemOpen
  \bibfield  {author} {\bibinfo {author} {\bibfnamefont {V.}~\bibnamefont
  {Lahtinen}}\ and\ \bibinfo {author} {\bibfnamefont {J.~K.}\ \bibnamefont
  {Pachos}},\ }\href@noop {} {\bibfield  {journal} {\bibinfo  {journal}
  {SciPost Phys.}\ }\textbf {\bibinfo {volume} {3}},\ \bibinfo {pages} {021}
  (\bibinfo {year} {2017})}\BibitemShut {NoStop}%
\bibitem [{\citenamefont {Wei}\ \emph {et~al.}(2019)\citenamefont {Wei},
  \citenamefont {Manna}, \citenamefont {Eich}, \citenamefont {Lee},\ and\
  \citenamefont {Moodera}}]{wei2019superconductivity}%
  \BibitemOpen
  \bibfield  {author} {\bibinfo {author} {\bibfnamefont {P.}~\bibnamefont
  {Wei}}, \bibinfo {author} {\bibfnamefont {S.}~\bibnamefont {Manna}}, \bibinfo
  {author} {\bibfnamefont {M.}~\bibnamefont {Eich}}, \bibinfo {author}
  {\bibfnamefont {P.}~\bibnamefont {Lee}}, \ and\ \bibinfo {author}
  {\bibfnamefont {J.}~\bibnamefont {Moodera}},\ }\href@noop {} {\bibfield
  {journal} {\bibinfo  {journal} {Physical Review Letters}\ }\textbf {\bibinfo
  {volume} {122}},\ \bibinfo {pages} {247002} (\bibinfo {year}
  {2019})}\BibitemShut {NoStop}%
\bibitem [{\citenamefont {Manna}\ \emph {et~al.}(2020)\citenamefont {Manna},
  \citenamefont {Wei}, \citenamefont {Xie}, \citenamefont {Law}, \citenamefont
  {Lee},\ and\ \citenamefont {Moodera}}]{manna2020signature}%
  \BibitemOpen
  \bibfield  {author} {\bibinfo {author} {\bibfnamefont {S.}~\bibnamefont
  {Manna}}, \bibinfo {author} {\bibfnamefont {P.}~\bibnamefont {Wei}}, \bibinfo
  {author} {\bibfnamefont {Y.}~\bibnamefont {Xie}}, \bibinfo {author}
  {\bibfnamefont {K.~T.}\ \bibnamefont {Law}}, \bibinfo {author} {\bibfnamefont
  {P.~A.}\ \bibnamefont {Lee}}, \ and\ \bibinfo {author} {\bibfnamefont
  {J.~S.}\ \bibnamefont {Moodera}},\ }\href@noop {} {\bibfield  {journal}
  {\bibinfo  {journal} {Proceedings of the National Academy of Sciences}\
  }\textbf {\bibinfo {volume} {117}},\ \bibinfo {pages} {8775} (\bibinfo {year}
  {2020})}\BibitemShut {NoStop}%
\bibitem [{\citenamefont {Sau}\ \emph {et~al.}(2010)\citenamefont {Sau},
  \citenamefont {Lutchyn}, \citenamefont {Tewari},\ and\ \citenamefont
  {Sarma}}]{sau2010generic}%
  \BibitemOpen
  \bibfield  {author} {\bibinfo {author} {\bibfnamefont {J.~D.}\ \bibnamefont
  {Sau}}, \bibinfo {author} {\bibfnamefont {R.~M.}\ \bibnamefont {Lutchyn}},
  \bibinfo {author} {\bibfnamefont {S.}~\bibnamefont {Tewari}}, \ and\ \bibinfo
  {author} {\bibfnamefont {S.~D.}\ \bibnamefont {Sarma}},\ }\href@noop {}
  {\bibfield  {journal} {\bibinfo  {journal} {Physical review letters}\
  }\textbf {\bibinfo {volume} {104}},\ \bibinfo {pages} {040502} (\bibinfo
  {year} {2010})}\BibitemShut {NoStop}%
\bibitem [{\citenamefont {Nadj-Perge}\ \emph {et~al.}(2013)\citenamefont
  {Nadj-Perge}, \citenamefont {Drozdov}, \citenamefont {Bernevig},\ and\
  \citenamefont {Yazdani}}]{nadj2013proposal}%
  \BibitemOpen
  \bibfield  {author} {\bibinfo {author} {\bibfnamefont {S.}~\bibnamefont
  {Nadj-Perge}}, \bibinfo {author} {\bibfnamefont {I.}~\bibnamefont {Drozdov}},
  \bibinfo {author} {\bibfnamefont {B.~A.}\ \bibnamefont {Bernevig}}, \ and\
  \bibinfo {author} {\bibfnamefont {A.}~\bibnamefont {Yazdani}},\ }\href@noop
  {} {\bibfield  {journal} {\bibinfo  {journal} {Physical Review B}\ }\textbf
  {\bibinfo {volume} {88}},\ \bibinfo {pages} {020407} (\bibinfo {year}
  {2013})}\BibitemShut {NoStop}%
\bibitem [{\citenamefont {Peng}\ \emph {et~al.}(2015)\citenamefont {Peng},
  \citenamefont {Pientka}, \citenamefont {Glazman},\ and\ \citenamefont
  {Von~Oppen}}]{peng2015strong}%
  \BibitemOpen
  \bibfield  {author} {\bibinfo {author} {\bibfnamefont {Y.}~\bibnamefont
  {Peng}}, \bibinfo {author} {\bibfnamefont {F.}~\bibnamefont {Pientka}},
  \bibinfo {author} {\bibfnamefont {L.~I.}\ \bibnamefont {Glazman}}, \ and\
  \bibinfo {author} {\bibfnamefont {F.}~\bibnamefont {Von~Oppen}},\ }\href@noop
  {} {\bibfield  {journal} {\bibinfo  {journal} {Physical review letters}\
  }\textbf {\bibinfo {volume} {114}},\ \bibinfo {pages} {106801} (\bibinfo
  {year} {2015})}\BibitemShut {NoStop}%
\bibitem [{\citenamefont {Brydon}\ \emph {et~al.}(2015)\citenamefont {Brydon},
  \citenamefont {Sarma}, \citenamefont {Hui},\ and\ \citenamefont
  {Sau}}]{brydon2015topological}%
  \BibitemOpen
  \bibfield  {author} {\bibinfo {author} {\bibfnamefont {P.}~\bibnamefont
  {Brydon}}, \bibinfo {author} {\bibfnamefont {S.~D.}\ \bibnamefont {Sarma}},
  \bibinfo {author} {\bibfnamefont {H.-Y.}\ \bibnamefont {Hui}}, \ and\
  \bibinfo {author} {\bibfnamefont {J.~D.}\ \bibnamefont {Sau}},\ }\href@noop
  {} {\bibfield  {journal} {\bibinfo  {journal} {Physical Review B}\ }\textbf
  {\bibinfo {volume} {91}},\ \bibinfo {pages} {064505} (\bibinfo {year}
  {2015})}\BibitemShut {NoStop}%
\bibitem [{\citenamefont {Cai}\ \emph {et~al.}(2010)\citenamefont {Cai},
  \citenamefont {Ruffieux}, \citenamefont {Jaafar}, \citenamefont {Bieri},
  \citenamefont {Braun}, \citenamefont {Blankenburg}, \citenamefont {Muoth},
  \citenamefont {Seitsonen}, \citenamefont {Saleh}, \citenamefont {Feng} \emph
  {et~al.}}]{cai2010atomically}%
  \BibitemOpen
  \bibfield  {author} {\bibinfo {author} {\bibfnamefont {J.}~\bibnamefont
  {Cai}}, \bibinfo {author} {\bibfnamefont {P.}~\bibnamefont {Ruffieux}},
  \bibinfo {author} {\bibfnamefont {R.}~\bibnamefont {Jaafar}}, \bibinfo
  {author} {\bibfnamefont {M.}~\bibnamefont {Bieri}}, \bibinfo {author}
  {\bibfnamefont {T.}~\bibnamefont {Braun}}, \bibinfo {author} {\bibfnamefont
  {S.}~\bibnamefont {Blankenburg}}, \bibinfo {author} {\bibfnamefont
  {M.}~\bibnamefont {Muoth}}, \bibinfo {author} {\bibfnamefont {A.~P.}\
  \bibnamefont {Seitsonen}}, \bibinfo {author} {\bibfnamefont {M.}~\bibnamefont
  {Saleh}}, \bibinfo {author} {\bibfnamefont {X.}~\bibnamefont {Feng}},  \emph
  {et~al.},\ }\href@noop {} {\bibfield  {journal} {\bibinfo  {journal}
  {Nature}\ }\textbf {\bibinfo {volume} {466}},\ \bibinfo {pages} {470}
  (\bibinfo {year} {2010})}\BibitemShut {NoStop}%
\bibitem [{\citenamefont {Ruffieux}\ \emph {et~al.}(2016)\citenamefont
  {Ruffieux}, \citenamefont {Wang}, \citenamefont {Yang}, \citenamefont
  {S{\'a}nchez-S{\'a}nchez}, \citenamefont {Liu}, \citenamefont {Dienel},
  \citenamefont {Talirz}, \citenamefont {Shinde}, \citenamefont {Pignedoli},
  \citenamefont {Passerone} \emph {et~al.}}]{ruffieux2016surface}%
  \BibitemOpen
  \bibfield  {author} {\bibinfo {author} {\bibfnamefont {P.}~\bibnamefont
  {Ruffieux}}, \bibinfo {author} {\bibfnamefont {S.}~\bibnamefont {Wang}},
  \bibinfo {author} {\bibfnamefont {B.}~\bibnamefont {Yang}}, \bibinfo {author}
  {\bibfnamefont {C.}~\bibnamefont {S{\'a}nchez-S{\'a}nchez}}, \bibinfo
  {author} {\bibfnamefont {J.}~\bibnamefont {Liu}}, \bibinfo {author}
  {\bibfnamefont {T.}~\bibnamefont {Dienel}}, \bibinfo {author} {\bibfnamefont
  {L.}~\bibnamefont {Talirz}}, \bibinfo {author} {\bibfnamefont
  {P.}~\bibnamefont {Shinde}}, \bibinfo {author} {\bibfnamefont {C.~A.}\
  \bibnamefont {Pignedoli}}, \bibinfo {author} {\bibfnamefont {D.}~\bibnamefont
  {Passerone}},  \emph {et~al.},\ }\href@noop {} {\bibfield  {journal}
  {\bibinfo  {journal} {Nature}\ }\textbf {\bibinfo {volume} {531}},\ \bibinfo
  {pages} {489} (\bibinfo {year} {2016})}\BibitemShut {NoStop}%
\bibitem [{\citenamefont {Liu}\ \emph {et~al.}(2023)\citenamefont {Liu},
  \citenamefont {Pawlak}, \citenamefont {Wang}, \citenamefont {Chen},
  \citenamefont {D’Astolfo}, \citenamefont {Drechsel}, \citenamefont {Zhou},
  \citenamefont {Häner}, \citenamefont {Decurtins}, \citenamefont {Aschauer},
  \citenamefont {Liu}, \citenamefont {Wulfhekel},\ and\ \citenamefont
  {Meyer}}]{liu_proximity-induced_2023}%
  \BibitemOpen
  \bibfield  {author} {\bibinfo {author} {\bibfnamefont {J.-C.}\ \bibnamefont
  {Liu}}, \bibinfo {author} {\bibfnamefont {R.}~\bibnamefont {Pawlak}},
  \bibinfo {author} {\bibfnamefont {X.}~\bibnamefont {Wang}}, \bibinfo {author}
  {\bibfnamefont {H.}~\bibnamefont {Chen}}, \bibinfo {author} {\bibfnamefont
  {P.}~\bibnamefont {D’Astolfo}}, \bibinfo {author} {\bibfnamefont
  {C.}~\bibnamefont {Drechsel}}, \bibinfo {author} {\bibfnamefont
  {P.}~\bibnamefont {Zhou}}, \bibinfo {author} {\bibfnamefont {R.}~\bibnamefont
  {Häner}}, \bibinfo {author} {\bibfnamefont {S.}~\bibnamefont {Decurtins}},
  \bibinfo {author} {\bibfnamefont {U.}~\bibnamefont {Aschauer}}, \bibinfo
  {author} {\bibfnamefont {S.-X.}\ \bibnamefont {Liu}}, \bibinfo {author}
  {\bibfnamefont {W.}~\bibnamefont {Wulfhekel}}, \ and\ \bibinfo {author}
  {\bibfnamefont {E.}~\bibnamefont {Meyer}},\ }\href@noop {} {\bibfield
  {journal} {\bibinfo  {journal} {ACS Materials Letters}\ }\textbf {\bibinfo
  {volume} {5}},\ \bibinfo {pages} {1083} (\bibinfo {year} {2023})},\ \bibinfo
  {note} {publisher: American Chemical Society}\BibitemShut {NoStop}%
\bibitem [{\citenamefont {Valla}\ \emph {et~al.}(1994)\citenamefont {Valla},
  \citenamefont {Pervan},\ and\ \citenamefont
  {Milun}}]{valla1994photoelectron}%
  \BibitemOpen
  \bibfield  {author} {\bibinfo {author} {\bibfnamefont {T.}~\bibnamefont
  {Valla}}, \bibinfo {author} {\bibfnamefont {P.}~\bibnamefont {Pervan}}, \
  and\ \bibinfo {author} {\bibfnamefont {M.}~\bibnamefont {Milun}},\
  }\href@noop {} {\bibfield  {journal} {\bibinfo  {journal} {Surface science}\
  }\textbf {\bibinfo {volume} {307}},\ \bibinfo {pages} {576} (\bibinfo {year}
  {1994})}\BibitemShut {NoStop}%
\bibitem [{\citenamefont {Lykhach}\ \emph {et~al.}(2003)\citenamefont
  {Lykhach}, \citenamefont {Pl{\v{s}}ek}, \citenamefont {Spirovov{\'a}},\ and\
  \citenamefont {Bastl}}]{lykhach2003thermal}%
  \BibitemOpen
  \bibfield  {author} {\bibinfo {author} {\bibfnamefont {Y.}~\bibnamefont
  {Lykhach}}, \bibinfo {author} {\bibfnamefont {J.}~\bibnamefont
  {Pl{\v{s}}ek}}, \bibinfo {author} {\bibfnamefont {I.}~\bibnamefont
  {Spirovov{\'a}}}, \ and\ \bibinfo {author} {\bibfnamefont {Z.}~\bibnamefont
  {Bastl}},\ }\href@noop {} {\bibfield  {journal} {\bibinfo  {journal}
  {Collection of Czechoslovak chemical communications}\ }\textbf {\bibinfo
  {volume} {68}},\ \bibinfo {pages} {1791} (\bibinfo {year}
  {2003})}\BibitemShut {NoStop}%
\bibitem [{\citenamefont {H{\"u}ger}\ \emph {et~al.}(2005)\citenamefont
  {H{\"u}ger}, \citenamefont {Wormeester},\ and\ \citenamefont
  {Osuch}}]{huger2005subsurface}%
  \BibitemOpen
  \bibfield  {author} {\bibinfo {author} {\bibfnamefont {E.}~\bibnamefont
  {H{\"u}ger}}, \bibinfo {author} {\bibfnamefont {H.}~\bibnamefont
  {Wormeester}}, \ and\ \bibinfo {author} {\bibfnamefont {K.}~\bibnamefont
  {Osuch}},\ }\href@noop {} {\bibfield  {journal} {\bibinfo  {journal} {Surface
  science}\ }\textbf {\bibinfo {volume} {580}},\ \bibinfo {pages} {173}
  (\bibinfo {year} {2005})}\BibitemShut {NoStop}%
\bibitem [{\citenamefont {Naftel}\ \emph {et~al.}(1999)\citenamefont {Naftel},
  \citenamefont {Bzowski},\ and\ \citenamefont {Sham}}]{naftel_study_1999}%
  \BibitemOpen
  \bibfield  {author} {\bibinfo {author} {\bibfnamefont {S.~J.}\ \bibnamefont
  {Naftel}}, \bibinfo {author} {\bibfnamefont {A.}~\bibnamefont {Bzowski}}, \
  and\ \bibinfo {author} {\bibfnamefont {T.~K.}\ \bibnamefont {Sham}},\
  }\href@noop {} {\bibfield  {journal} {\bibinfo  {journal} {Journal of Alloys
  and Compounds}\ ,\ \bibinfo {pages} {7}} (\bibinfo {year}
  {1999})}\BibitemShut {NoStop}%
\bibitem [{\citenamefont {Ruckman}\ and\ \citenamefont
  {Jiang}(1988)}]{ruckman_growth_1988}%
  \BibitemOpen
  \bibfield  {author} {\bibinfo {author} {\bibfnamefont {M.~W.}\ \bibnamefont
  {Ruckman}}\ and\ \bibinfo {author} {\bibfnamefont {L.-Q.}\ \bibnamefont
  {Jiang}},\ }\href@noop {} {\bibfield  {journal} {\bibinfo  {journal}
  {Physical Review B}\ }\textbf {\bibinfo {volume} {38}},\ \bibinfo {pages}
  {2959} (\bibinfo {year} {1988})}\BibitemShut {NoStop}%
\bibitem [{\citenamefont {Wei}\ \emph {et~al.}(2016)\citenamefont {Wei},
  \citenamefont {Katmis}, \citenamefont {Chang},\ and\ \citenamefont
  {Moodera}}]{wei2016induced}%
  \BibitemOpen
  \bibfield  {author} {\bibinfo {author} {\bibfnamefont {P.}~\bibnamefont
  {Wei}}, \bibinfo {author} {\bibfnamefont {F.}~\bibnamefont {Katmis}},
  \bibinfo {author} {\bibfnamefont {C.-Z.}\ \bibnamefont {Chang}}, \ and\
  \bibinfo {author} {\bibfnamefont {J.~S.}\ \bibnamefont {Moodera}},\
  }\href@noop {} {\bibfield  {journal} {\bibinfo  {journal} {Nano letters}\
  }\textbf {\bibinfo {volume} {16}},\ \bibinfo {pages} {2714} (\bibinfo {year}
  {2016})}\BibitemShut {NoStop}%
\bibitem [{\citenamefont {Beck}\ \emph {et~al.}(2023)\citenamefont {Beck},
  \citenamefont {Ny{\'a}ri}, \citenamefont {Schneider}, \citenamefont
  {R{\'o}zsa}, \citenamefont {L{\'a}szl{\'o}ffy}, \citenamefont {Palot{\'a}s},
  \citenamefont {Szunyogh}, \citenamefont {Ujfalussy}, \citenamefont {Wiebe},\
  and\ \citenamefont {Wiesendanger}}]{beck2023search}%
  \BibitemOpen
  \bibfield  {author} {\bibinfo {author} {\bibfnamefont {P.}~\bibnamefont
  {Beck}}, \bibinfo {author} {\bibfnamefont {B.}~\bibnamefont {Ny{\'a}ri}},
  \bibinfo {author} {\bibfnamefont {L.}~\bibnamefont {Schneider}}, \bibinfo
  {author} {\bibfnamefont {L.}~\bibnamefont {R{\'o}zsa}}, \bibinfo {author}
  {\bibfnamefont {A.}~\bibnamefont {L{\'a}szl{\'o}ffy}}, \bibinfo {author}
  {\bibfnamefont {K.}~\bibnamefont {Palot{\'a}s}}, \bibinfo {author}
  {\bibfnamefont {L.}~\bibnamefont {Szunyogh}}, \bibinfo {author}
  {\bibfnamefont {B.}~\bibnamefont {Ujfalussy}}, \bibinfo {author}
  {\bibfnamefont {J.}~\bibnamefont {Wiebe}}, \ and\ \bibinfo {author}
  {\bibfnamefont {R.}~\bibnamefont {Wiesendanger}},\ }\href@noop {} {\bibfield
  {journal} {\bibinfo  {journal} {Communications Physics}\ }\textbf {\bibinfo
  {volume} {6}},\ \bibinfo {pages} {83} (\bibinfo {year} {2023})}\BibitemShut
  {NoStop}%
\bibitem [{\citenamefont {Yu}(1965)}]{yu1965bound}%
  \BibitemOpen
  \bibfield  {author} {\bibinfo {author} {\bibfnamefont {L.}~\bibnamefont
  {Yu}},\ }\href@noop {} {\bibfield  {journal} {\bibinfo  {journal} {Wu Li
  Hsueh Pao (China) Supersedes Chung-Kuo Wu Li Hsueh For English translation
  see Chin. J. Phys.(Peking)(Engl. Transl.)}\ }\textbf {\bibinfo {volume} {21}}
  (\bibinfo {year} {1965})}\BibitemShut {NoStop}%
\bibitem [{\citenamefont {Shiba}(1968)}]{shiba1968classical}%
  \BibitemOpen
  \bibfield  {author} {\bibinfo {author} {\bibfnamefont {H.}~\bibnamefont
  {Shiba}},\ }\href@noop {} {\bibfield  {journal} {\bibinfo  {journal}
  {Progress of theoretical Physics}\ }\textbf {\bibinfo {volume} {40}},\
  \bibinfo {pages} {435} (\bibinfo {year} {1968})}\BibitemShut {NoStop}%
\bibitem [{\citenamefont {Rusinov}(1969)}]{rusinov1969theory}%
  \BibitemOpen
  \bibfield  {author} {\bibinfo {author} {\bibfnamefont {A.}~\bibnamefont
  {Rusinov}},\ }\href@noop {} {\bibfield  {journal} {\bibinfo  {journal} {Sov.
  Phys. JETP}\ }\textbf {\bibinfo {volume} {29}},\ \bibinfo {pages} {1101}
  (\bibinfo {year} {1969})}\BibitemShut {NoStop}%
\bibitem [{\citenamefont {Kneisel}\ \emph {et~al.}(2003)\citenamefont {Kneisel}
  \emph {et~al.}}]{kneisel2003surface}%
  \BibitemOpen
  \bibfield  {author} {\bibinfo {author} {\bibfnamefont {P.}~\bibnamefont
  {Kneisel}} \emph {et~al.},\ }in\ \href@noop {} {\emph {\bibinfo {booktitle}
  {Proc. of the 11th Workshop on RF Superconductivity,
  L{\"u}beck/Travem{\"u}nde, Germany}}}\ (\bibinfo {year} {2003})\BibitemShut
  {NoStop}%
\bibitem [{\citenamefont {Shen}\ \emph {et~al.}(2007)\citenamefont {Shen},
  \citenamefont {Ma}, \citenamefont {Lee},\ and\ \citenamefont
  {Zaera}}]{shen_oxygen_2007}%
  \BibitemOpen
  \bibfield  {author} {\bibinfo {author} {\bibfnamefont {M.}~\bibnamefont
  {Shen}}, \bibinfo {author} {\bibfnamefont {Q.}~\bibnamefont {Ma}}, \bibinfo
  {author} {\bibfnamefont {I.}~\bibnamefont {Lee}}, \ and\ \bibinfo {author}
  {\bibfnamefont {F.}~\bibnamefont {Zaera}},\ }\href@noop {} {\bibfield
  {journal} {\bibinfo  {journal} {The Journal of Physical Chemistry C}\
  }\textbf {\bibinfo {volume} {111}},\ \bibinfo {pages} {6033} (\bibinfo {year}
  {2007})}\BibitemShut {NoStop}%
\bibitem [{\citenamefont {Odobesko}\ \emph {et~al.}(2019)\citenamefont
  {Odobesko}, \citenamefont {Haldar}, \citenamefont {Wilfert}, \citenamefont
  {Hagen}, \citenamefont {Jung}, \citenamefont {Schmidt}, \citenamefont
  {Sessi}, \citenamefont {Vogt}, \citenamefont {Heinze},\ and\ \citenamefont
  {Bode}}]{odobesko2019preparation}%
  \BibitemOpen
  \bibfield  {author} {\bibinfo {author} {\bibfnamefont {A.~B.}\ \bibnamefont
  {Odobesko}}, \bibinfo {author} {\bibfnamefont {S.}~\bibnamefont {Haldar}},
  \bibinfo {author} {\bibfnamefont {S.}~\bibnamefont {Wilfert}}, \bibinfo
  {author} {\bibfnamefont {J.}~\bibnamefont {Hagen}}, \bibinfo {author}
  {\bibfnamefont {J.}~\bibnamefont {Jung}}, \bibinfo {author} {\bibfnamefont
  {N.}~\bibnamefont {Schmidt}}, \bibinfo {author} {\bibfnamefont
  {P.}~\bibnamefont {Sessi}}, \bibinfo {author} {\bibfnamefont
  {M.}~\bibnamefont {Vogt}}, \bibinfo {author} {\bibfnamefont {S.}~\bibnamefont
  {Heinze}}, \ and\ \bibinfo {author} {\bibfnamefont {M.}~\bibnamefont
  {Bode}},\ }\href@noop {} {\bibfield  {journal} {\bibinfo  {journal} {Physical
  Review B}\ }\textbf {\bibinfo {volume} {99}},\ \bibinfo {pages} {115437}
  (\bibinfo {year} {2019})}\BibitemShut {NoStop}%
\bibitem [{\citenamefont {Koller}\ \emph {et~al.}(2001)\citenamefont {Koller},
  \citenamefont {Bergermayer}, \citenamefont {Kresse}, \citenamefont
  {Hebenstreit}, \citenamefont {Konvicka}, \citenamefont {Schmid},
  \citenamefont {Podloucky},\ and\ \citenamefont
  {Varga}}]{koller2001structure}%
  \BibitemOpen
  \bibfield  {author} {\bibinfo {author} {\bibfnamefont {R.}~\bibnamefont
  {Koller}}, \bibinfo {author} {\bibfnamefont {W.}~\bibnamefont {Bergermayer}},
  \bibinfo {author} {\bibfnamefont {G.}~\bibnamefont {Kresse}}, \bibinfo
  {author} {\bibfnamefont {E.}~\bibnamefont {Hebenstreit}}, \bibinfo {author}
  {\bibfnamefont {C.}~\bibnamefont {Konvicka}}, \bibinfo {author}
  {\bibfnamefont {M.}~\bibnamefont {Schmid}}, \bibinfo {author} {\bibfnamefont
  {R.}~\bibnamefont {Podloucky}}, \ and\ \bibinfo {author} {\bibfnamefont
  {P.}~\bibnamefont {Varga}},\ }\href@noop {} {\bibfield  {journal} {\bibinfo
  {journal} {Surface science}\ }\textbf {\bibinfo {volume} {480}},\ \bibinfo
  {pages} {11} (\bibinfo {year} {2001})}\BibitemShut {NoStop}%
\bibitem [{\citenamefont {Schneider}\ \emph {et~al.}(2023)\citenamefont
  {Schneider}, \citenamefont {Ton}, \citenamefont {Ioannidis}, \citenamefont
  {Neuhaus-Steinmetz}, \citenamefont {Posske}, \citenamefont {Wiesendanger},\
  and\ \citenamefont {Wiebe}}]{schneider_proximity_2023}%
  \BibitemOpen
  \bibfield  {author} {\bibinfo {author} {\bibfnamefont {L.}~\bibnamefont
  {Schneider}}, \bibinfo {author} {\bibfnamefont {K.~T.}\ \bibnamefont {Ton}},
  \bibinfo {author} {\bibfnamefont {I.}~\bibnamefont {Ioannidis}}, \bibinfo
  {author} {\bibfnamefont {J.}~\bibnamefont {Neuhaus-Steinmetz}}, \bibinfo
  {author} {\bibfnamefont {T.}~\bibnamefont {Posske}}, \bibinfo {author}
  {\bibfnamefont {R.}~\bibnamefont {Wiesendanger}}, \ and\ \bibinfo {author}
  {\bibfnamefont {J.}~\bibnamefont {Wiebe}},\ }\href@noop {} {\bibfield
  {journal} {\bibinfo  {journal} {Nature}\ }\textbf {\bibinfo {volume} {621}},\
  \bibinfo {pages} {60} (\bibinfo {year} {2023})}\BibitemShut {NoStop}%
\bibitem [{\citenamefont {Valla}\ \emph {et~al.}(1995)\citenamefont {Valla},
  \citenamefont {Pervan},\ and\ \citenamefont
  {Milun}}]{valla_interaction_1995}%
  \BibitemOpen
  \bibfield  {author} {\bibinfo {author} {\bibfnamefont {T.}~\bibnamefont
  {Valla}}, \bibinfo {author} {\bibfnamefont {P.}~\bibnamefont {Pervan}}, \
  and\ \bibinfo {author} {\bibfnamefont {M.}~\bibnamefont {Milun}},\
  }\href@noop {} {\bibfield  {journal} {\bibinfo  {journal} {Applied Surface
  Science}\ }\textbf {\bibinfo {volume} {89}},\ \bibinfo {pages} {375}
  (\bibinfo {year} {1995})}\BibitemShut {NoStop}%
\bibitem [{\citenamefont {Tempas}\ \emph {et~al.}(2018)\citenamefont {Tempas},
  \citenamefont {Morris}, \citenamefont {Wisman}, \citenamefont {Le},
  \citenamefont {Din}, \citenamefont {Williams}, \citenamefont {Wang},
  \citenamefont {Polezhaev}, \citenamefont {Rahman}, \citenamefont {Caulton},\
  and\ \citenamefont {Tait}}]{tempas_2018_redox}%
  \BibitemOpen
  \bibfield  {author} {\bibinfo {author} {\bibfnamefont {C.~D.}\ \bibnamefont
  {Tempas}}, \bibinfo {author} {\bibfnamefont {T.}~\bibnamefont {Morris}},
  \bibinfo {author} {\bibfnamefont {D.~L.}\ \bibnamefont {Wisman}}, \bibinfo
  {author} {\bibfnamefont {D.}~\bibnamefont {Le}}, \bibinfo {author}
  {\bibfnamefont {N.~U.}\ \bibnamefont {Din}}, \bibinfo {author} {\bibfnamefont
  {C.~G.}\ \bibnamefont {Williams}}, \bibinfo {author} {\bibfnamefont
  {M.}~\bibnamefont {Wang}}, \bibinfo {author} {\bibfnamefont {A.~V.}\
  \bibnamefont {Polezhaev}}, \bibinfo {author} {\bibfnamefont {T.~S.}\
  \bibnamefont {Rahman}}, \bibinfo {author} {\bibfnamefont {K.~G.}\
  \bibnamefont {Caulton}}, \ and\ \bibinfo {author} {\bibfnamefont {S.~L.}\
  \bibnamefont {Tait}},\ }\href@noop {} {\bibfield  {journal} {\bibinfo
  {journal} {Chem. Sci.}\ }\textbf {\bibinfo {volume} {9}},\ \bibinfo {pages}
  {1674} (\bibinfo {year} {2018})}\BibitemShut {NoStop}%
\bibitem [{\citenamefont {Huang}\ \emph {et~al.}(2020)\citenamefont {Huang},
  \citenamefont {Drost}, \citenamefont {Senkpiel}, \citenamefont {Padurariu},
  \citenamefont {Kubala}, \citenamefont {Yeyati}, \citenamefont {Cuevas},
  \citenamefont {Ankerhold}, \citenamefont {Kern},\ and\ \citenamefont
  {Ast}}]{huang2020quantum}%
  \BibitemOpen
  \bibfield  {author} {\bibinfo {author} {\bibfnamefont {H.}~\bibnamefont
  {Huang}}, \bibinfo {author} {\bibfnamefont {R.}~\bibnamefont {Drost}},
  \bibinfo {author} {\bibfnamefont {J.}~\bibnamefont {Senkpiel}}, \bibinfo
  {author} {\bibfnamefont {C.}~\bibnamefont {Padurariu}}, \bibinfo {author}
  {\bibfnamefont {B.}~\bibnamefont {Kubala}}, \bibinfo {author} {\bibfnamefont
  {A.~L.}\ \bibnamefont {Yeyati}}, \bibinfo {author} {\bibfnamefont {J.~C.}\
  \bibnamefont {Cuevas}}, \bibinfo {author} {\bibfnamefont {J.}~\bibnamefont
  {Ankerhold}}, \bibinfo {author} {\bibfnamefont {K.}~\bibnamefont {Kern}}, \
  and\ \bibinfo {author} {\bibfnamefont {C.~R.}\ \bibnamefont {Ast}},\
  }\href@noop {} {\bibfield  {journal} {\bibinfo  {journal} {Communications
  Physics}\ }\textbf {\bibinfo {volume} {3}},\ \bibinfo {pages} {1} (\bibinfo
  {year} {2020})}\BibitemShut {NoStop}%
\bibitem [{\citenamefont {Kralj}\ \emph {et~al.}(2003)\citenamefont {Kralj},
  \citenamefont {Pervan}, \citenamefont {Milun}, \citenamefont {Wandelt},
  \citenamefont {Mandrino},\ and\ \citenamefont {Jenko}}]{kralj_hraes_2003}%
  \BibitemOpen
  \bibfield  {author} {\bibinfo {author} {\bibfnamefont {M.}~\bibnamefont
  {Kralj}}, \bibinfo {author} {\bibfnamefont {P.}~\bibnamefont {Pervan}},
  \bibinfo {author} {\bibfnamefont {M.}~\bibnamefont {Milun}}, \bibinfo
  {author} {\bibfnamefont {K.}~\bibnamefont {Wandelt}}, \bibinfo {author}
  {\bibfnamefont {D.}~\bibnamefont {Mandrino}}, \ and\ \bibinfo {author}
  {\bibfnamefont {M.}~\bibnamefont {Jenko}},\ }\href@noop {} {\bibfield
  {journal} {\bibinfo  {journal} {Surface Science}\ ,\ \bibinfo {pages} {11}}
  (\bibinfo {year} {2003})}\BibitemShut {NoStop}%
\bibitem [{\citenamefont {Kralj}\ \emph {et~al.}(1999)\citenamefont {Kralj},
  \citenamefont {Pervan}, \citenamefont {Milun}, \citenamefont {Schneider},
  \citenamefont {Schaefer}, \citenamefont {Rosenhahn},\ and\ \citenamefont
  {Wandelt}}]{Kralj_1999_stm}%
  \BibitemOpen
  \bibfield  {author} {\bibinfo {author} {\bibfnamefont {M.}~\bibnamefont
  {Kralj}}, \bibinfo {author} {\bibfnamefont {P.}~\bibnamefont {Pervan}},
  \bibinfo {author} {\bibfnamefont {M.}~\bibnamefont {Milun}}, \bibinfo
  {author} {\bibfnamefont {J.}~\bibnamefont {Schneider}}, \bibinfo {author}
  {\bibfnamefont {B.}~\bibnamefont {Schaefer}}, \bibinfo {author}
  {\bibfnamefont {A.}~\bibnamefont {Rosenhahn}}, \ and\ \bibinfo {author}
  {\bibfnamefont {K.}~\bibnamefont {Wandelt}},\ }\href@noop {} {\bibfield
  {journal} {\bibinfo  {journal} {Fizika A}\ ,\ \bibinfo {pages} {123}}
  (\bibinfo {year} {1999})}\BibitemShut {NoStop}%
\bibitem [{\citenamefont {Yoshimoto}\ \emph {et~al.}(2012)\citenamefont
  {Yoshimoto}, \citenamefont {Kim}, \citenamefont {Sato}, \citenamefont
  {Inukai},\ and\ \citenamefont {Itaya}}]{yoshimoto2012potential}%
  \BibitemOpen
  \bibfield  {author} {\bibinfo {author} {\bibfnamefont {S.}~\bibnamefont
  {Yoshimoto}}, \bibinfo {author} {\bibfnamefont {Y.-G.}\ \bibnamefont {Kim}},
  \bibinfo {author} {\bibfnamefont {K.}~\bibnamefont {Sato}}, \bibinfo {author}
  {\bibfnamefont {J.}~\bibnamefont {Inukai}}, \ and\ \bibinfo {author}
  {\bibfnamefont {K.}~\bibnamefont {Itaya}},\ }\href@noop {} {\bibfield
  {journal} {\bibinfo  {journal} {Physical Chemistry Chemical Physics}\
  }\textbf {\bibinfo {volume} {14}},\ \bibinfo {pages} {2286} (\bibinfo {year}
  {2012})}\BibitemShut {NoStop}%
\bibitem [{\citenamefont {Hammer}\ \emph {et~al.}(2014)\citenamefont {Hammer},
  \citenamefont {Sander}, \citenamefont {F{\"o}rster}, \citenamefont {Kiel},
  \citenamefont {Meinel},\ and\ \citenamefont {Widdra}}]{hammer2014surface}%
  \BibitemOpen
  \bibfield  {author} {\bibinfo {author} {\bibfnamefont {R.}~\bibnamefont
  {Hammer}}, \bibinfo {author} {\bibfnamefont {A.}~\bibnamefont {Sander}},
  \bibinfo {author} {\bibfnamefont {S.}~\bibnamefont {F{\"o}rster}}, \bibinfo
  {author} {\bibfnamefont {M.}~\bibnamefont {Kiel}}, \bibinfo {author}
  {\bibfnamefont {K.}~\bibnamefont {Meinel}}, \ and\ \bibinfo {author}
  {\bibfnamefont {W.}~\bibnamefont {Widdra}},\ }\href@noop {} {\bibfield
  {journal} {\bibinfo  {journal} {Physical Review B}\ }\textbf {\bibinfo
  {volume} {90}},\ \bibinfo {pages} {035446} (\bibinfo {year}
  {2014})}\BibitemShut {NoStop}%
\bibitem [{\citenamefont {Trembu{\l}owicz}\ \emph {et~al.}(2021)\citenamefont
  {Trembu{\l}owicz}, \citenamefont {Sabik},\ and\ \citenamefont
  {Grodzicki}}]{trembulowicz2021100}%
  \BibitemOpen
  \bibfield  {author} {\bibinfo {author} {\bibfnamefont {A.}~\bibnamefont
  {Trembu{\l}owicz}}, \bibinfo {author} {\bibfnamefont {A.}~\bibnamefont
  {Sabik}}, \ and\ \bibinfo {author} {\bibfnamefont {M.}~\bibnamefont
  {Grodzicki}},\ }\href@noop {} {\bibfield  {journal} {\bibinfo  {journal}
  {Molecules}\ }\textbf {\bibinfo {volume} {26}},\ \bibinfo {pages} {2393}
  (\bibinfo {year} {2021})}\BibitemShut {NoStop}%
\bibitem [{\citenamefont {Tomanic}\ \emph {et~al.}(2016)\citenamefont
  {Tomanic}, \citenamefont {Schackert}, \citenamefont {Wulfhekel},
  \citenamefont {S\"urgers},\ and\ \citenamefont
  {L\"ohneysen}}]{tomanic_2016_two-band}%
  \BibitemOpen
  \bibfield  {author} {\bibinfo {author} {\bibfnamefont {T.}~\bibnamefont
  {Tomanic}}, \bibinfo {author} {\bibfnamefont {M.}~\bibnamefont {Schackert}},
  \bibinfo {author} {\bibfnamefont {W.}~\bibnamefont {Wulfhekel}}, \bibinfo
  {author} {\bibfnamefont {C.}~\bibnamefont {S\"urgers}}, \ and\ \bibinfo
  {author} {\bibfnamefont {H.~v.}\ \bibnamefont {L\"ohneysen}},\ }\href@noop {}
  {\bibfield  {journal} {\bibinfo  {journal} {Phys. Rev. B}\ }\textbf {\bibinfo
  {volume} {94}},\ \bibinfo {pages} {220503} (\bibinfo {year}
  {2016})}\BibitemShut {NoStop}%
\bibitem [{\citenamefont {Zhussupbekov}\ \emph {et~al.}(2020)\citenamefont
  {Zhussupbekov}, \citenamefont {Walshe}, \citenamefont {Bozhko}, \citenamefont
  {Ionov}, \citenamefont {Fleischer}, \citenamefont {Norton}, \citenamefont
  {Zhussupbekova}, \citenamefont {Semenov}, \citenamefont {Shvets},\ and\
  \citenamefont {Walls}}]{zhussupbekov2020oxidation}%
  \BibitemOpen
  \bibfield  {author} {\bibinfo {author} {\bibfnamefont {K.}~\bibnamefont
  {Zhussupbekov}}, \bibinfo {author} {\bibfnamefont {K.}~\bibnamefont
  {Walshe}}, \bibinfo {author} {\bibfnamefont {S.~I.}\ \bibnamefont {Bozhko}},
  \bibinfo {author} {\bibfnamefont {A.}~\bibnamefont {Ionov}}, \bibinfo
  {author} {\bibfnamefont {K.}~\bibnamefont {Fleischer}}, \bibinfo {author}
  {\bibfnamefont {E.}~\bibnamefont {Norton}}, \bibinfo {author} {\bibfnamefont
  {A.}~\bibnamefont {Zhussupbekova}}, \bibinfo {author} {\bibfnamefont
  {V.}~\bibnamefont {Semenov}}, \bibinfo {author} {\bibfnamefont {I.~V.}\
  \bibnamefont {Shvets}}, \ and\ \bibinfo {author} {\bibfnamefont
  {B.}~\bibnamefont {Walls}},\ }\href@noop {} {\bibfield  {journal} {\bibinfo
  {journal} {Scientific reports}\ }\textbf {\bibinfo {volume} {10}},\ \bibinfo
  {pages} {1} (\bibinfo {year} {2020})}\BibitemShut {NoStop}%
\bibitem [{\citenamefont {Razinkin}\ and\ \citenamefont
  {Kuznetsov}(2010)}]{razinkin2010scanning}%
  \BibitemOpen
  \bibfield  {author} {\bibinfo {author} {\bibfnamefont {A.}~\bibnamefont
  {Razinkin}}\ and\ \bibinfo {author} {\bibfnamefont {M.}~\bibnamefont
  {Kuznetsov}},\ }\href@noop {} {\bibfield  {journal} {\bibinfo  {journal} {The
  Physics of Metals and Metallography}\ }\textbf {\bibinfo {volume} {110}},\
  \bibinfo {pages} {531} (\bibinfo {year} {2010})}\BibitemShut {NoStop}%
\bibitem [{\citenamefont {Arfaoui}\ \emph {et~al.}(2004)\citenamefont
  {Arfaoui}, \citenamefont {Cousty},\ and\ \citenamefont
  {Guillot}}]{arfaoui2004model}%
  \BibitemOpen
  \bibfield  {author} {\bibinfo {author} {\bibfnamefont {I.}~\bibnamefont
  {Arfaoui}}, \bibinfo {author} {\bibfnamefont {J.}~\bibnamefont {Cousty}}, \
  and\ \bibinfo {author} {\bibfnamefont {C.}~\bibnamefont {Guillot}},\
  }\href@noop {} {\bibfield  {journal} {\bibinfo  {journal} {Surface science}\
  }\textbf {\bibinfo {volume} {557}},\ \bibinfo {pages} {119} (\bibinfo {year}
  {2004})}\BibitemShut {NoStop}%
\bibitem [{\citenamefont {Berman}\ \emph {et~al.}(2023)\citenamefont {Berman},
  \citenamefont {Zhussupbekova}, \citenamefont {Walls}, \citenamefont {Walshe},
  \citenamefont {Bozhko}, \citenamefont {Ionov}, \citenamefont {O'Regan},
  \citenamefont {Shvets},\ and\ \citenamefont
  {Zhussupbekov}}]{berman2023unraveling}%
  \BibitemOpen
  \bibfield  {author} {\bibinfo {author} {\bibfnamefont {S.}~\bibnamefont
  {Berman}}, \bibinfo {author} {\bibfnamefont {A.}~\bibnamefont
  {Zhussupbekova}}, \bibinfo {author} {\bibfnamefont {B.}~\bibnamefont
  {Walls}}, \bibinfo {author} {\bibfnamefont {K.}~\bibnamefont {Walshe}},
  \bibinfo {author} {\bibfnamefont {S.~I.}\ \bibnamefont {Bozhko}}, \bibinfo
  {author} {\bibfnamefont {A.}~\bibnamefont {Ionov}}, \bibinfo {author}
  {\bibfnamefont {D.~D.}\ \bibnamefont {O'Regan}}, \bibinfo {author}
  {\bibfnamefont {I.~V.}\ \bibnamefont {Shvets}}, \ and\ \bibinfo {author}
  {\bibfnamefont {K.}~\bibnamefont {Zhussupbekov}},\ }\href@noop {} {\bibfield
  {journal} {\bibinfo  {journal} {Physical Review B}\ }\textbf {\bibinfo
  {volume} {107}},\ \bibinfo {pages} {165425} (\bibinfo {year}
  {2023})}\BibitemShut {NoStop}%
\bibitem [{\citenamefont {Levi}\ and\ \citenamefont
  {Kotrla}(1996)}]{Theory_levi_1996}%
  \BibitemOpen
  \bibfield  {author} {\bibinfo {author} {\bibfnamefont {A.}~\bibnamefont
  {Levi}}\ and\ \bibinfo {author} {\bibfnamefont {M.}~\bibnamefont {Kotrla}},\
  }\href@noop {} {\bibfield  {journal} {\bibinfo  {journal} {Journal of Physics
  Condensed Matter}\ }\textbf {\bibinfo {volume} {9}},\ \bibinfo {pages} {299}
  (\bibinfo {year} {1996})}\BibitemShut {NoStop}%
\bibitem [{\citenamefont {Feenstra}\ and\ \citenamefont
  {Hla}(2015)}]{LandoltBornstein2015:sm_lbs_978-3-662-47736-6_2}%
  \BibitemOpen
  \bibfield  {author} {\bibinfo {author} {\bibfnamefont {R.~M.}\ \bibnamefont
  {Feenstra}}\ and\ \bibinfo {author} {\bibfnamefont {S.~W.}\ \bibnamefont
  {Hla}},\ }\href@noop {} {\enquote {\bibinfo {title} {2.1 introduction to
  scanning tunneling microscopy of metals and semiconductor},}\ } (\bibinfo
  {year} {2015})\BibitemShut {NoStop}%
\bibitem [{\citenamefont {Veit}\ \emph {et~al.}(2019)\citenamefont {Veit},
  \citenamefont {Kautz}, \citenamefont {Farber},\ and\ \citenamefont
  {Sibener}}]{veit2019oxygen}%
  \BibitemOpen
  \bibfield  {author} {\bibinfo {author} {\bibfnamefont {R.~D.}\ \bibnamefont
  {Veit}}, \bibinfo {author} {\bibfnamefont {N.~A.}\ \bibnamefont {Kautz}},
  \bibinfo {author} {\bibfnamefont {R.~G.}\ \bibnamefont {Farber}}, \ and\
  \bibinfo {author} {\bibfnamefont {S.}~\bibnamefont {Sibener}},\ }\href@noop
  {} {\bibfield  {journal} {\bibinfo  {journal} {Surface Science}\ }\textbf
  {\bibinfo {volume} {688}},\ \bibinfo {pages} {63} (\bibinfo {year}
  {2019})}\BibitemShut {NoStop}%
\bibitem [{\citenamefont {An}\ \emph {et~al.}(2003)\citenamefont {An},
  \citenamefont {Fukuyama}, \citenamefont {Yokogawa},\ and\ \citenamefont
  {Yoshimura}}]{an2003surface}%
  \BibitemOpen
  \bibfield  {author} {\bibinfo {author} {\bibfnamefont {B.}~\bibnamefont
  {An}}, \bibinfo {author} {\bibfnamefont {S.}~\bibnamefont {Fukuyama}},
  \bibinfo {author} {\bibfnamefont {K.}~\bibnamefont {Yokogawa}}, \ and\
  \bibinfo {author} {\bibfnamefont {M.}~\bibnamefont {Yoshimura}},\ }\href@noop
  {} {\bibfield  {journal} {\bibinfo  {journal} {Physical Review B}\ }\textbf
  {\bibinfo {volume} {68}},\ \bibinfo {pages} {115423} (\bibinfo {year}
  {2003})}\BibitemShut {NoStop}%
\bibitem [{\citenamefont {Wulfhekel}\ \emph {et~al.}(2000)\citenamefont
  {Wulfhekel}, \citenamefont {Zavaliche}, \citenamefont {Porrati},
  \citenamefont {Oepen},\ and\ \citenamefont {Kirschner}}]{wulfhekel2000nano}%
  \BibitemOpen
  \bibfield  {author} {\bibinfo {author} {\bibfnamefont {W.}~\bibnamefont
  {Wulfhekel}}, \bibinfo {author} {\bibfnamefont {F.}~\bibnamefont
  {Zavaliche}}, \bibinfo {author} {\bibfnamefont {F.}~\bibnamefont {Porrati}},
  \bibinfo {author} {\bibfnamefont {H.}~\bibnamefont {Oepen}}, \ and\ \bibinfo
  {author} {\bibfnamefont {J.}~\bibnamefont {Kirschner}},\ }\href@noop {}
  {\bibfield  {journal} {\bibinfo  {journal} {Europhysics Letters}\ }\textbf
  {\bibinfo {volume} {49}},\ \bibinfo {pages} {651} (\bibinfo {year}
  {2000})}\BibitemShut {NoStop}%
\bibitem [{\citenamefont {Von~Bergmann}\ \emph {et~al.}(2004)\citenamefont
  {Von~Bergmann}, \citenamefont {Bode},\ and\ \citenamefont
  {Wiesendanger}}]{von2004magnetism}%
  \BibitemOpen
  \bibfield  {author} {\bibinfo {author} {\bibfnamefont {K.}~\bibnamefont
  {Von~Bergmann}}, \bibinfo {author} {\bibfnamefont {M.}~\bibnamefont {Bode}},
  \ and\ \bibinfo {author} {\bibfnamefont {R.}~\bibnamefont {Wiesendanger}},\
  }\href@noop {} {\bibfield  {journal} {\bibinfo  {journal} {Physical Review
  B}\ }\textbf {\bibinfo {volume} {70}},\ \bibinfo {pages} {174455} (\bibinfo
  {year} {2004})}\BibitemShut {NoStop}%
\bibitem [{\citenamefont {Nadj-Perge}\ \emph {et~al.}(2014)\citenamefont
  {Nadj-Perge}, \citenamefont {Drozdov}, \citenamefont {Li}, \citenamefont
  {Chen}, \citenamefont {Jeon}, \citenamefont {Seo}, \citenamefont {MacDonald},
  \citenamefont {Bernevig},\ and\ \citenamefont
  {Yazdani}}]{nadj2014observation}%
  \BibitemOpen
  \bibfield  {author} {\bibinfo {author} {\bibfnamefont {S.}~\bibnamefont
  {Nadj-Perge}}, \bibinfo {author} {\bibfnamefont {I.~K.}\ \bibnamefont
  {Drozdov}}, \bibinfo {author} {\bibfnamefont {J.}~\bibnamefont {Li}},
  \bibinfo {author} {\bibfnamefont {H.}~\bibnamefont {Chen}}, \bibinfo {author}
  {\bibfnamefont {S.}~\bibnamefont {Jeon}}, \bibinfo {author} {\bibfnamefont
  {J.}~\bibnamefont {Seo}}, \bibinfo {author} {\bibfnamefont {A.~H.}\
  \bibnamefont {MacDonald}}, \bibinfo {author} {\bibfnamefont {B.~A.}\
  \bibnamefont {Bernevig}}, \ and\ \bibinfo {author} {\bibfnamefont
  {A.}~\bibnamefont {Yazdani}},\ }\href@noop {} {\bibfield  {journal} {\bibinfo
   {journal} {Science}\ }\textbf {\bibinfo {volume} {346}},\ \bibinfo {pages}
  {602} (\bibinfo {year} {2014})}\BibitemShut {NoStop}%
\bibitem [{\citenamefont {Anderson}(1959)}]{Anderson1959}%
  \BibitemOpen
  \bibfield  {author} {\bibinfo {author} {\bibfnamefont {W.}~\bibnamefont
  {Anderson}},\ }\href@noop {} {\bibfield  {journal} {\bibinfo  {journal} {J.
  Phys. Chem. Solids}\ }\textbf {\bibinfo {volume} {11}},\ \bibinfo {pages}
  {26} (\bibinfo {year} {1959})}\BibitemShut {NoStop}%
\bibitem [{\citenamefont {Ji}\ \emph {et~al.}(2008)\citenamefont {Ji},
  \citenamefont {Zhang}, \citenamefont {Fu}, \citenamefont {Chen},
  \citenamefont {Ma}, \citenamefont {Li}, \citenamefont {Duan}, \citenamefont
  {Jia},\ and\ \citenamefont {Xue}}]{ji2008high}%
  \BibitemOpen
  \bibfield  {author} {\bibinfo {author} {\bibfnamefont {S.-H.}\ \bibnamefont
  {Ji}}, \bibinfo {author} {\bibfnamefont {T.}~\bibnamefont {Zhang}}, \bibinfo
  {author} {\bibfnamefont {Y.-S.}\ \bibnamefont {Fu}}, \bibinfo {author}
  {\bibfnamefont {X.}~\bibnamefont {Chen}}, \bibinfo {author} {\bibfnamefont
  {X.-C.}\ \bibnamefont {Ma}}, \bibinfo {author} {\bibfnamefont
  {J.}~\bibnamefont {Li}}, \bibinfo {author} {\bibfnamefont {W.-H.}\
  \bibnamefont {Duan}}, \bibinfo {author} {\bibfnamefont {J.-F.}\ \bibnamefont
  {Jia}}, \ and\ \bibinfo {author} {\bibfnamefont {Q.-K.}\ \bibnamefont
  {Xue}},\ }\href@noop {} {\bibfield  {journal} {\bibinfo  {journal} {Physical
  review letters}\ }\textbf {\bibinfo {volume} {100}},\ \bibinfo {pages}
  {226801} (\bibinfo {year} {2008})}\BibitemShut {NoStop}%
\bibitem [{\citenamefont {Franke}\ \emph {et~al.}(2011)\citenamefont {Franke},
  \citenamefont {Schulze},\ and\ \citenamefont {Pascual}}]{Franke11}%
  \BibitemOpen
  \bibfield  {author} {\bibinfo {author} {\bibfnamefont {K.~J.}\ \bibnamefont
  {Franke}}, \bibinfo {author} {\bibfnamefont {G.}~\bibnamefont {Schulze}}, \
  and\ \bibinfo {author} {\bibfnamefont {J.~I.}\ \bibnamefont {Pascual}},\
  }\href@noop {} {\bibfield  {journal} {\bibinfo  {journal} {Science}\ }\textbf
  {\bibinfo {volume} {332}},\ \bibinfo {pages} {940} (\bibinfo {year}
  {2011})}\BibitemShut {NoStop}%
\bibitem [{\citenamefont {Arnold}(1978)}]{arnold1978theory}%
  \BibitemOpen
  \bibfield  {author} {\bibinfo {author} {\bibfnamefont {G.~B.}\ \bibnamefont
  {Arnold}},\ }\href@noop {} {\bibfield  {journal} {\bibinfo  {journal}
  {Physical Review B}\ }\textbf {\bibinfo {volume} {18}},\ \bibinfo {pages}
  {1076} (\bibinfo {year} {1978})}\BibitemShut {NoStop}%
\bibitem [{\citenamefont {De~Gennes}\ and\ \citenamefont
  {Saint-James}(1963)}]{de1963elementary}%
  \BibitemOpen
  \bibfield  {author} {\bibinfo {author} {\bibfnamefont {P.}~\bibnamefont
  {De~Gennes}}\ and\ \bibinfo {author} {\bibfnamefont {D.}~\bibnamefont
  {Saint-James}},\ }\href@noop {} {\bibfield  {journal} {\bibinfo  {journal}
  {Phys. Letters}\ }\textbf {\bibinfo {volume} {4}} (\bibinfo {year}
  {1963})}\BibitemShut {NoStop}%
\bibitem [{\citenamefont {Ortuzar}\ \emph {et~al.}(2023)\citenamefont
  {Ortuzar}, \citenamefont {Pascual}, \citenamefont {Bergeret},\ and\
  \citenamefont {Cazalilla}}]{ortuzar2023theory}%
  \BibitemOpen
  \bibfield  {author} {\bibinfo {author} {\bibfnamefont {J.}~\bibnamefont
  {Ortuzar}}, \bibinfo {author} {\bibfnamefont {J.~I.}\ \bibnamefont
  {Pascual}}, \bibinfo {author} {\bibfnamefont {F.~S.}\ \bibnamefont
  {Bergeret}}, \ and\ \bibinfo {author} {\bibfnamefont {M.~A.}\ \bibnamefont
  {Cazalilla}},\ }\href@noop {} {\bibfield  {journal} {\bibinfo  {journal}
  {Physical Review B}\ }\textbf {\bibinfo {volume} {108}},\ \bibinfo {pages}
  {024511} (\bibinfo {year} {2023})}\BibitemShut {NoStop}%
\bibitem [{\citenamefont {Trivini}\ \emph {et~al.}(2023)\citenamefont
  {Trivini}, \citenamefont {Ortuzar}, \citenamefont {Vaxevani}, \citenamefont
  {Li}, \citenamefont {Bergeret}, \citenamefont {Cazalilla},\ and\
  \citenamefont {Pascual}}]{trivini2023cooper}%
  \BibitemOpen
  \bibfield  {author} {\bibinfo {author} {\bibfnamefont {S.}~\bibnamefont
  {Trivini}}, \bibinfo {author} {\bibfnamefont {J.}~\bibnamefont {Ortuzar}},
  \bibinfo {author} {\bibfnamefont {K.}~\bibnamefont {Vaxevani}}, \bibinfo
  {author} {\bibfnamefont {J.}~\bibnamefont {Li}}, \bibinfo {author}
  {\bibfnamefont {F.~S.}\ \bibnamefont {Bergeret}}, \bibinfo {author}
  {\bibfnamefont {M.~A.}\ \bibnamefont {Cazalilla}}, \ and\ \bibinfo {author}
  {\bibfnamefont {J.~I.}\ \bibnamefont {Pascual}},\ }\href@noop {} {\bibfield
  {journal} {\bibinfo  {journal} {Physical Review Letters}\ }\textbf {\bibinfo
  {volume} {130}},\ \bibinfo {pages} {136004} (\bibinfo {year}
  {2023})}\BibitemShut {NoStop}%
\bibitem [{Sup()}]{SuppMat}%
  \BibitemOpen
  \href@noop {} {}\bibinfo {note} {See Supplemental Material at [URL] for
  further information on the determination of the superconducting gap of the Au
  film using a superconducting tip in Fig.4 and 5; additional XPS spectra and
  analysis on carbon; images of as-deposite Au and low temperature annealing;
  models describing the superconductors' surface mophology. The Supplemental
  Material also contains Refs. [39,43,51,52,56].}\BibitemShut {Stop}%
\bibitem [{\citenamefont {Heinrich}\ \emph
  {et~al.}(2013{\natexlab{a}})\citenamefont {Heinrich}, \citenamefont {Braun},
  \citenamefont {Pascual},\ and\ \citenamefont {Franke}}]{Heinrich13a}%
  \BibitemOpen
  \bibfield  {author} {\bibinfo {author} {\bibfnamefont {B.~W.}\ \bibnamefont
  {Heinrich}}, \bibinfo {author} {\bibfnamefont {L.}~\bibnamefont {Braun}},
  \bibinfo {author} {\bibfnamefont {J.~I.}\ \bibnamefont {Pascual}}, \ and\
  \bibinfo {author} {\bibfnamefont {K.~J.}\ \bibnamefont {Franke}},\
  }\href@noop {} {\bibfield  {journal} {\bibinfo  {journal} {Nature Phys.}\
  }\textbf {\bibinfo {volume} {9}},\ \bibinfo {pages} {765} (\bibinfo {year}
  {2013}{\natexlab{a}})}\BibitemShut {NoStop}%
\bibitem [{\citenamefont {Heinrich}\ \emph
  {et~al.}(2013{\natexlab{b}})\citenamefont {Heinrich}, \citenamefont {Ahmadi},
  \citenamefont {M\"uller}, \citenamefont {Braun}, \citenamefont {Pascual},\
  and\ \citenamefont {Franke}}]{Heinrich13b}%
  \BibitemOpen
  \bibfield  {author} {\bibinfo {author} {\bibfnamefont {B.~W.}\ \bibnamefont
  {Heinrich}}, \bibinfo {author} {\bibfnamefont {G.}~\bibnamefont {Ahmadi}},
  \bibinfo {author} {\bibfnamefont {V.~L.}\ \bibnamefont {M\"uller}}, \bibinfo
  {author} {\bibfnamefont {L.}~\bibnamefont {Braun}}, \bibinfo {author}
  {\bibfnamefont {J.~I.}\ \bibnamefont {Pascual}}, \ and\ \bibinfo {author}
  {\bibfnamefont {K.~J.}\ \bibnamefont {Franke}},\ }\href@noop {} {\bibfield
  {journal} {\bibinfo  {journal} {Nano Lett.}\ }\textbf {\bibinfo {volume}
  {13}},\ \bibinfo {pages} {4840} (\bibinfo {year}
  {2013}{\natexlab{b}})}\BibitemShut {NoStop}%
\bibitem [{\citenamefont {Heinrich}\ \emph {et~al.}(2015)\citenamefont
  {Heinrich}, \citenamefont {Braun}, \citenamefont {Pascual},\ and\
  \citenamefont {Franke}}]{Heinrich15}%
  \BibitemOpen
  \bibfield  {author} {\bibinfo {author} {\bibfnamefont {B.~W.}\ \bibnamefont
  {Heinrich}}, \bibinfo {author} {\bibfnamefont {L.}~\bibnamefont {Braun}},
  \bibinfo {author} {\bibfnamefont {J.~I.}\ \bibnamefont {Pascual}}, \ and\
  \bibinfo {author} {\bibfnamefont {K.~J.}\ \bibnamefont {Franke}},\
  }\href@noop {} {\bibfield  {journal} {\bibinfo  {journal} {Nano Lett.}\
  }\textbf {\bibinfo {volume} {15}},\ \bibinfo {pages} {4024} (\bibinfo {year}
  {2015})}\BibitemShut {NoStop}%
\bibitem [{\citenamefont {Rubio-Verd\'u}\ \emph {et~al.}(2018)\citenamefont
  {Rubio-Verd\'u}, \citenamefont {Sarasola}, \citenamefont {Choi},
  \citenamefont {Majzik}, \citenamefont {Ebeling}, \citenamefont {Calvo},
  \citenamefont {Ugeda}, \citenamefont {Garcia-Lekue}, \citenamefont
  {S\'anchez-Portal},\ and\ \citenamefont {Pascual}}]{Rubio-Verdu18}%
  \BibitemOpen
  \bibfield  {author} {\bibinfo {author} {\bibfnamefont {C.}~\bibnamefont
  {Rubio-Verd\'u}}, \bibinfo {author} {\bibfnamefont {A.}~\bibnamefont
  {Sarasola}}, \bibinfo {author} {\bibfnamefont {D.-J.}\ \bibnamefont {Choi}},
  \bibinfo {author} {\bibfnamefont {Z.}~\bibnamefont {Majzik}}, \bibinfo
  {author} {\bibfnamefont {R.}~\bibnamefont {Ebeling}}, \bibinfo {author}
  {\bibfnamefont {M.~R.}\ \bibnamefont {Calvo}}, \bibinfo {author}
  {\bibfnamefont {M.~M.}\ \bibnamefont {Ugeda}}, \bibinfo {author}
  {\bibfnamefont {A.}~\bibnamefont {Garcia-Lekue}}, \bibinfo {author}
  {\bibfnamefont {D.}~\bibnamefont {S\'anchez-Portal}}, \ and\ \bibinfo
  {author} {\bibfnamefont {J.~I.}\ \bibnamefont {Pascual}},\ }\href@noop {}
  {\bibfield  {journal} {\bibinfo  {journal} {Commun. Phys}\ }\textbf {\bibinfo
  {volume} {1}},\ \bibinfo {pages} {15} (\bibinfo {year} {2018})}\BibitemShut
  {NoStop}%
\bibitem [{\citenamefont {Farinacci}\ \emph {et~al.}(2020)\citenamefont
  {Farinacci}, \citenamefont {Ahmadi}, \citenamefont {Ruby}, \citenamefont
  {Reecht}, \citenamefont {Heinrich}, \citenamefont {Czekelius}, \citenamefont
  {{von Oppen}},\ and\ \citenamefont {Franke}}]{Farinacci20}%
  \BibitemOpen
  \bibfield  {author} {\bibinfo {author} {\bibfnamefont {L.}~\bibnamefont
  {Farinacci}}, \bibinfo {author} {\bibfnamefont {G.}~\bibnamefont {Ahmadi}},
  \bibinfo {author} {\bibfnamefont {M.}~\bibnamefont {Ruby}}, \bibinfo {author}
  {\bibfnamefont {G.}~\bibnamefont {Reecht}}, \bibinfo {author} {\bibfnamefont
  {B.~W.}\ \bibnamefont {Heinrich}}, \bibinfo {author} {\bibfnamefont
  {C.}~\bibnamefont {Czekelius}}, \bibinfo {author} {\bibfnamefont
  {F.}~\bibnamefont {{von Oppen}}}, \ and\ \bibinfo {author} {\bibfnamefont
  {K.~J.}\ \bibnamefont {Franke}},\ }\href@noop {} {\bibfield  {journal}
  {\bibinfo  {journal} {Phys. Rev. lett.}\ }\textbf {\bibinfo {volume} {125}},\
  \bibinfo {pages} {256805} (\bibinfo {year} {2020})}\BibitemShut {NoStop}%
\bibitem [{\citenamefont {Rubio-Verd\'u}\ \emph {et~al.}(2021)\citenamefont
  {Rubio-Verd\'u}, \citenamefont {Zald\'ivar}, \citenamefont {\v{Z}itko},\ and\
  \citenamefont {Pascual}}]{Rubio-Verdu21a}%
  \BibitemOpen
  \bibfield  {author} {\bibinfo {author} {\bibfnamefont {C.}~\bibnamefont
  {Rubio-Verd\'u}}, \bibinfo {author} {\bibfnamefont {J.}~\bibnamefont
  {Zald\'ivar}}, \bibinfo {author} {\bibfnamefont {R.}~\bibnamefont
  {\v{Z}itko}}, \ and\ \bibinfo {author} {\bibfnamefont {J.~I.}\ \bibnamefont
  {Pascual}},\ }\href@noop {} {\bibfield  {journal} {\bibinfo  {journal} {Phys.
  Rev. Lett.}\ }\textbf {\bibinfo {volume} {126}},\ \bibinfo {pages} {017001}
  (\bibinfo {year} {2021})}\BibitemShut {NoStop}%
\bibitem [{\citenamefont {Wang}\ \emph {et~al.}(2015)\citenamefont {Wang},
  \citenamefont {Pang}, \citenamefont {Kuang}, \citenamefont {Shi},
  \citenamefont {Shang}, \citenamefont {Liu},\ and\ \citenamefont
  {Lin}}]{wang2015intramolecularly}%
  \BibitemOpen
  \bibfield  {author} {\bibinfo {author} {\bibfnamefont {W.}~\bibnamefont
  {Wang}}, \bibinfo {author} {\bibfnamefont {R.}~\bibnamefont {Pang}}, \bibinfo
  {author} {\bibfnamefont {G.}~\bibnamefont {Kuang}}, \bibinfo {author}
  {\bibfnamefont {X.}~\bibnamefont {Shi}}, \bibinfo {author} {\bibfnamefont
  {X.}~\bibnamefont {Shang}}, \bibinfo {author} {\bibfnamefont {P.~N.}\
  \bibnamefont {Liu}}, \ and\ \bibinfo {author} {\bibfnamefont
  {N.}~\bibnamefont {Lin}},\ }\href@noop {} {\bibfield  {journal} {\bibinfo
  {journal} {Physical Review B}\ }\textbf {\bibinfo {volume} {91}},\ \bibinfo
  {pages} {045440} (\bibinfo {year} {2015})}\BibitemShut {NoStop}%
\bibitem [{\citenamefont {Liu}\ \emph {et~al.}(2017)\citenamefont {Liu},
  \citenamefont {Fu}, \citenamefont {Guan}, \citenamefont {Shao}, \citenamefont
  {Meng}, \citenamefont {Guo},\ and\ \citenamefont {Wang}}]{liu2017iron}%
  \BibitemOpen
  \bibfield  {author} {\bibinfo {author} {\bibfnamefont {B.}~\bibnamefont
  {Liu}}, \bibinfo {author} {\bibfnamefont {H.}~\bibnamefont {Fu}}, \bibinfo
  {author} {\bibfnamefont {J.}~\bibnamefont {Guan}}, \bibinfo {author}
  {\bibfnamefont {B.}~\bibnamefont {Shao}}, \bibinfo {author} {\bibfnamefont
  {S.}~\bibnamefont {Meng}}, \bibinfo {author} {\bibfnamefont {J.}~\bibnamefont
  {Guo}}, \ and\ \bibinfo {author} {\bibfnamefont {W.}~\bibnamefont {Wang}},\
  }\href@noop {} {\bibfield  {journal} {\bibinfo  {journal} {ACS nano}\
  }\textbf {\bibinfo {volume} {11}},\ \bibinfo {pages} {11402} (\bibinfo {year}
  {2017})}\BibitemShut {NoStop}%
\bibitem [{\citenamefont {Rolf}\ \emph {et~al.}(2018)\citenamefont {Rolf},
  \citenamefont {Lotze}, \citenamefont {Czekelius}, \citenamefont {Heinrich},\
  and\ \citenamefont {Franke}}]{rolf2018visualizing}%
  \BibitemOpen
  \bibfield  {author} {\bibinfo {author} {\bibfnamefont {D.}~\bibnamefont
  {Rolf}}, \bibinfo {author} {\bibfnamefont {C.}~\bibnamefont {Lotze}},
  \bibinfo {author} {\bibfnamefont {C.}~\bibnamefont {Czekelius}}, \bibinfo
  {author} {\bibfnamefont {B.~W.}\ \bibnamefont {Heinrich}}, \ and\ \bibinfo
  {author} {\bibfnamefont {K.~J.}\ \bibnamefont {Franke}},\ }\href@noop {}
  {\bibfield  {journal} {\bibinfo  {journal} {The journal of physical chemistry
  letters}\ }\textbf {\bibinfo {volume} {9}},\ \bibinfo {pages} {6563}
  (\bibinfo {year} {2018})}\BibitemShut {NoStop}%
\bibitem [{\citenamefont {Rolf}\ \emph {et~al.}(2019)\citenamefont {Rolf},
  \citenamefont {Maa{\ss}}, \citenamefont {Lotze}, \citenamefont {Czekelius},
  \citenamefont {Heinrich}, \citenamefont {Tegeder},\ and\ \citenamefont
  {Franke}}]{rolf2019correlation}%
  \BibitemOpen
  \bibfield  {author} {\bibinfo {author} {\bibfnamefont {D.}~\bibnamefont
  {Rolf}}, \bibinfo {author} {\bibfnamefont {F.}~\bibnamefont {Maa{\ss}}},
  \bibinfo {author} {\bibfnamefont {C.}~\bibnamefont {Lotze}}, \bibinfo
  {author} {\bibfnamefont {C.}~\bibnamefont {Czekelius}}, \bibinfo {author}
  {\bibfnamefont {B.~W.}\ \bibnamefont {Heinrich}}, \bibinfo {author}
  {\bibfnamefont {P.}~\bibnamefont {Tegeder}}, \ and\ \bibinfo {author}
  {\bibfnamefont {K.~J.}\ \bibnamefont {Franke}},\ }\href@noop {} {\bibfield
  {journal} {\bibinfo  {journal} {The Journal of Physical Chemistry C}\
  }\textbf {\bibinfo {volume} {123}},\ \bibinfo {pages} {7425} (\bibinfo {year}
  {2019})}\BibitemShut {NoStop}%
\bibitem [{\citenamefont {Mugarza}\ \emph {et~al.}(2010)\citenamefont
  {Mugarza}, \citenamefont {Lorente}, \citenamefont {Ordej{\'o}n},
  \citenamefont {Krull}, \citenamefont {Stepanow}, \citenamefont {Bocquet},
  \citenamefont {Fraxedas}, \citenamefont {Ceballos},\ and\ \citenamefont
  {Gambardella}}]{Mugarza10}%
  \BibitemOpen
  \bibfield  {author} {\bibinfo {author} {\bibfnamefont {A.}~\bibnamefont
  {Mugarza}}, \bibinfo {author} {\bibfnamefont {N.}~\bibnamefont {Lorente}},
  \bibinfo {author} {\bibfnamefont {P.}~\bibnamefont {Ordej{\'o}n}}, \bibinfo
  {author} {\bibfnamefont {C.}~\bibnamefont {Krull}}, \bibinfo {author}
  {\bibfnamefont {S.}~\bibnamefont {Stepanow}}, \bibinfo {author}
  {\bibfnamefont {M.-L.}\ \bibnamefont {Bocquet}}, \bibinfo {author}
  {\bibfnamefont {J.}~\bibnamefont {Fraxedas}}, \bibinfo {author}
  {\bibfnamefont {G.}~\bibnamefont {Ceballos}}, \ and\ \bibinfo {author}
  {\bibfnamefont {P.}~\bibnamefont {Gambardella}},\ }\href@noop {} {\bibfield
  {journal} {\bibinfo  {journal} {Phys. Rev. lett.}\ }\textbf {\bibinfo
  {volume} {105}},\ \bibinfo {pages} {115702} (\bibinfo {year}
  {2010})}\BibitemShut {NoStop}%
\bibitem [{\citenamefont {Farinacci}\ \emph {et~al.}(2018)\citenamefont
  {Farinacci}, \citenamefont {Ahmadi}, \citenamefont {Reecht}, \citenamefont
  {Ruby}, \citenamefont {Bogdanoff}, \citenamefont {Peters}, \citenamefont
  {Heinrich}, \citenamefont {von Oppen},\ and\ \citenamefont
  {Franke}}]{farinacci2018tuning}%
  \BibitemOpen
  \bibfield  {author} {\bibinfo {author} {\bibfnamefont {L.}~\bibnamefont
  {Farinacci}}, \bibinfo {author} {\bibfnamefont {G.}~\bibnamefont {Ahmadi}},
  \bibinfo {author} {\bibfnamefont {G.}~\bibnamefont {Reecht}}, \bibinfo
  {author} {\bibfnamefont {M.}~\bibnamefont {Ruby}}, \bibinfo {author}
  {\bibfnamefont {N.}~\bibnamefont {Bogdanoff}}, \bibinfo {author}
  {\bibfnamefont {O.}~\bibnamefont {Peters}}, \bibinfo {author} {\bibfnamefont
  {B.~W.}\ \bibnamefont {Heinrich}}, \bibinfo {author} {\bibfnamefont
  {F.}~\bibnamefont {von Oppen}}, \ and\ \bibinfo {author} {\bibfnamefont
  {K.~J.}\ \bibnamefont {Franke}},\ }\href@noop {} {\bibfield  {journal}
  {\bibinfo  {journal} {Physical review letters}\ }\textbf {\bibinfo {volume}
  {121}},\ \bibinfo {pages} {196803} (\bibinfo {year} {2018})}\BibitemShut
  {NoStop}%
\bibitem [{\citenamefont {Liu}\ \emph {et~al.}(2022)\citenamefont {Liu},
  \citenamefont {Li}, \citenamefont {Xue}, \citenamefont {Wang}, \citenamefont
  {Huang}, \citenamefont {Yang}, \citenamefont {Chen}, \citenamefont {Guan},
  \citenamefont {Li}, \citenamefont {Zheng} \emph {et~al.}}]{liu2022quantum}%
  \BibitemOpen
  \bibfield  {author} {\bibinfo {author} {\bibfnamefont {Y.}~\bibnamefont
  {Liu}}, \bibinfo {author} {\bibfnamefont {C.}~\bibnamefont {Li}}, \bibinfo
  {author} {\bibfnamefont {F.-H.}\ \bibnamefont {Xue}}, \bibinfo {author}
  {\bibfnamefont {Y.}~\bibnamefont {Wang}}, \bibinfo {author} {\bibfnamefont
  {H.}~\bibnamefont {Huang}}, \bibinfo {author} {\bibfnamefont
  {H.}~\bibnamefont {Yang}}, \bibinfo {author} {\bibfnamefont {J.}~\bibnamefont
  {Chen}}, \bibinfo {author} {\bibfnamefont {D.-D.}\ \bibnamefont {Guan}},
  \bibinfo {author} {\bibfnamefont {Y.-Y.}\ \bibnamefont {Li}}, \bibinfo
  {author} {\bibfnamefont {H.}~\bibnamefont {Zheng}},  \emph {et~al.},\
  }\href@noop {} {\bibfield  {journal} {\bibinfo  {journal} {arXiv preprint
  arXiv:2207.05313}\ } (\bibinfo {year} {2022})}\BibitemShut {NoStop}%
\end{thebibliography}
%

\end{document}